\documentclass[preprint,showpacs,amsmath,amssymb]{revtex4}

\usepackage{graphicx}% Include figure files
\usepackage{dcolumn}% Align table columns on decimal point
\usepackage{bm}% bold math

\begin{document}

\title{Magnetic Field Induced Non-Fermi-Liquid Behavior in YbAgGe Single Crystal}

\author{S. L. Bud'ko$^\ast$, E. Q. T. Morosan$^{\ast\dagger}$, and P. C. Canfield$^{\ast\dagger}$}
\affiliation{$^\ast$Ames Laboratory and $^\dagger$Department of
Physics and Astronomy, Iowa State University, Ames, IA 50011, USA}

\date{\today}

\begin{abstract}
Detailed anisotropic resistivity and heat capacity measurements
down to $\sim$ 0.4 K and up to 140 kOe are reported for a single
crystalline YbAgGe. Based on these data YbAgGe, a member of the
hexagonal RAgGe series, can be classified as new, {\it
stoichiometric} heavy fermion compound with two magnetic ordering
temperatures below 1 K and field-induced non-Fermi-liquid behavior
above 45-70 kOe and 80-110 kOe for $H
\parallel (ab)$ and $H \parallel c$ respectively.
\end{abstract}

\maketitle

% main text
\section{Introduction}
\begin{figure}
\begin{center}
\includegraphics[angle=0,width=120mm]{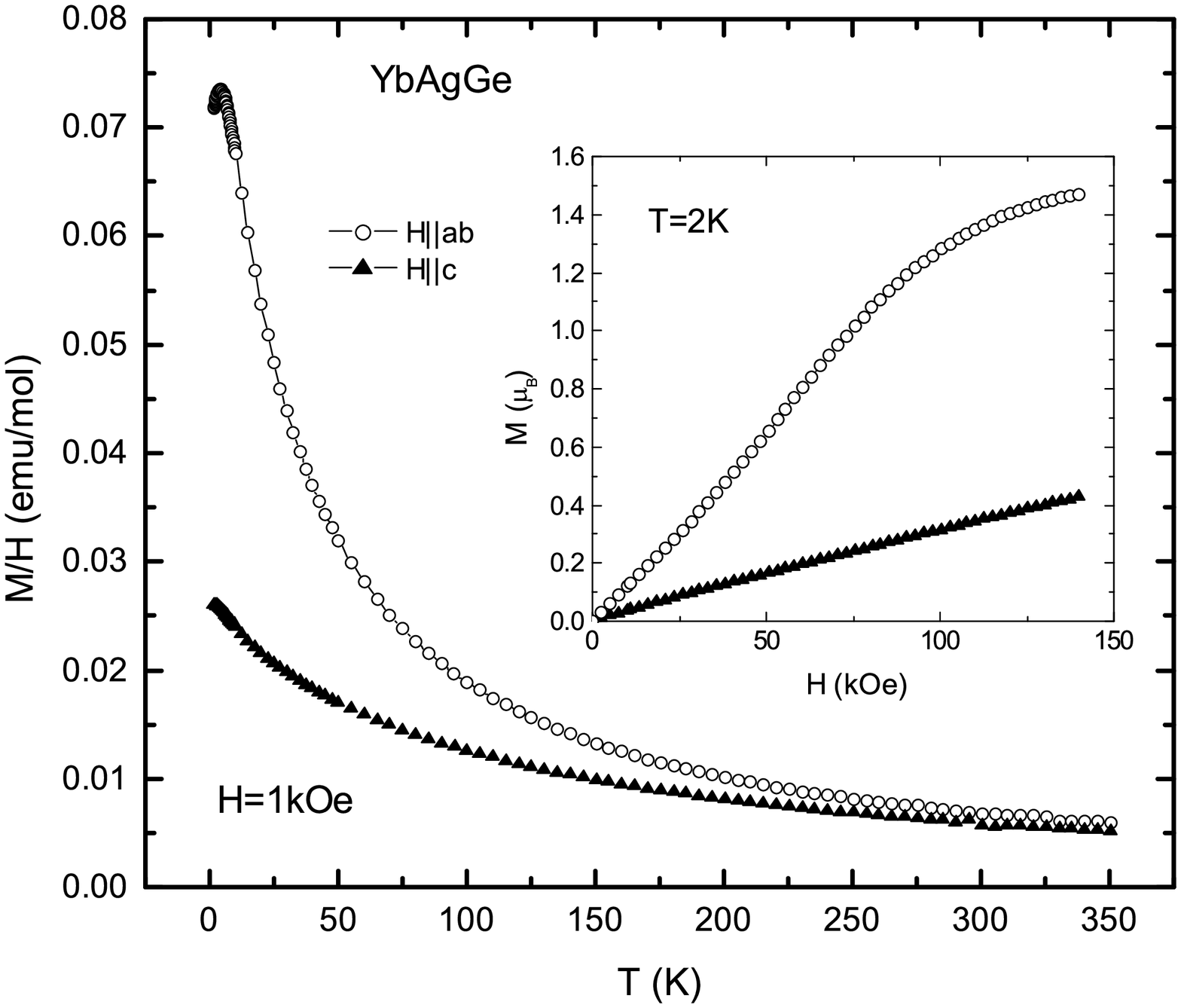}
\end{center}
\caption{Anisotropic temperature-dependent DC susceptibility and
(inset) field-dependent magnetization of YbAgGe.}\label{MTH}
\end{figure}

\begin{figure}
\begin{center}
\includegraphics[angle=0,width=120mm]{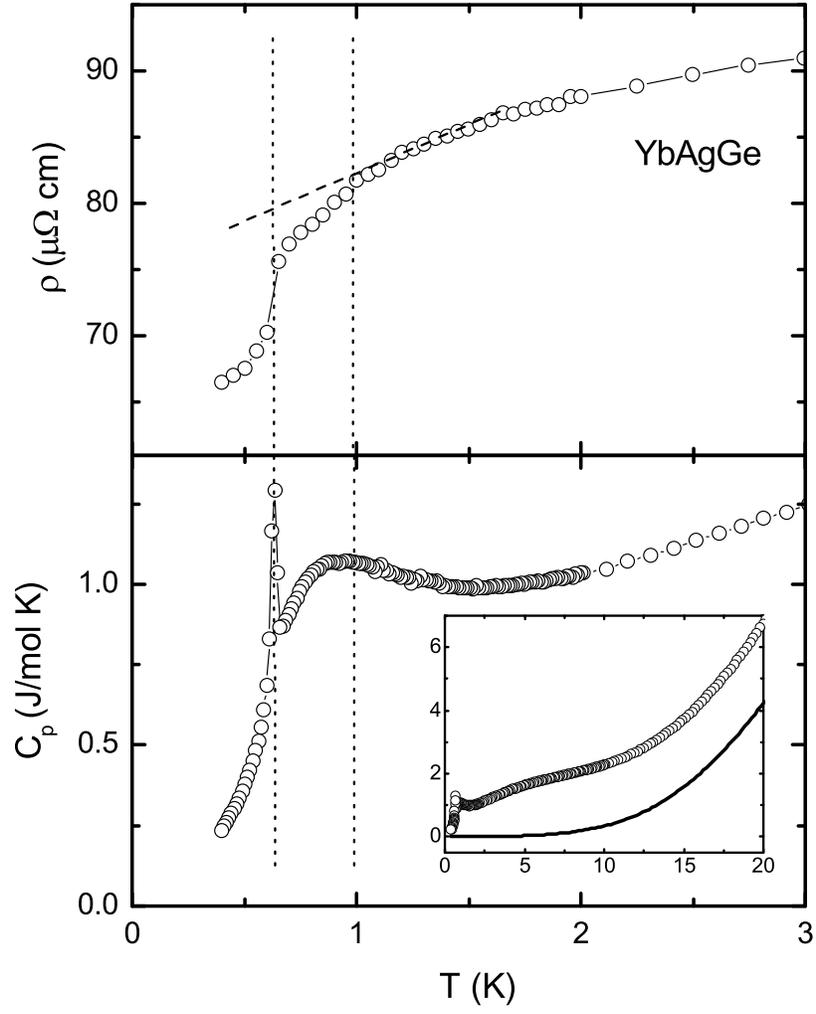}
\end{center}
\caption{Low temperature zero field resistivity (upper panel) and
specific heat (lower panel)of YbAgGe. Dotted lines point to
magnetic transitions. Inset: $C_p(T)$ of YbAgGe and LuAgGe (line)
up to 20 K.}\label{RTCT}
\end{figure}

YbAgGe is the penultimate member of the hexagonal RAgGe series
\cite{mor03a} and was recently identified
\cite{mor03a,bey98a,can03a,kat03a} as a new Yb-based heavy fermion
compound. Magnetization measurements on YbAgGe down to 1.8 K
\cite{mor03a} show moderate anisotropy (at low temperatures
$\chi_{ab}/\chi_c \approx$ 3) and a loss of local moment character
below $\sim 20$ K (Fig. \ref{MTH}). The in-plane $M(H)$ at $T =$ 2
K shows a trend toward saturation whereas $H
\parallel c$ field-dependent magnetization continues to be
virtually linear below 140 kOe (Fig. \ref{MTH}, inset). Initial
thermodynamic and transport measurements down to 0.4 K
\cite{bey98a,can03a,mor03a} reveal two magnetic transitions, a
higher one at $\approx$ 1 K, and a lower one, with very sharp
features in $\rho(T)$ and $C_p(T)$, at $\approx$ 0.65 K (Fig.
\ref{RTCT}). Given that the magnetic entropy inferred from the
Fig. \ref{RTCT} is only $\sim 5\%$ of $R\ln2$ at 1 K and only
reaches $R\ln2$ by $\sim$25 K it seems likely that these
transitions are associated with a small moment ordering. Based on
these measurements the compound was anticipated to be close to the
quantum critical point. The linear component of $C_p(T)$,
$\gamma$, is $\sim$150 mJ/mol K$^2$ between 12 K and 20 K.
$C_p(T)/T$ rises up to $\sim$1200 mJ/mol K$^2$ for $T \sim$ 1 K
but given the presence of the magnetic transitions below 1 K, it
is difficult to unambiguously evaluate the electronic specific
heat. Grossly speaking, 150 mJ/mol K$^2$ $< \gamma <$ 1 J/mol
K$^2$ leading to an estimate 10 K $< T_K <$ 100 K for the Kondo
temperature, $T_K$.

Since the number of the Yb-based heavy fermion compounds is
relatively small \cite{ste84a,fis92a,hew93a} any new member of the
family attracts attention \cite{fis92a,tho94a}. As an up to date
example, YbRh$_2$Si$_2$, a heavy fermion antiferromagnet
\cite{tro00a}, became a subject of intensive, rewarding
exploration \cite{geg02a,ish02a,cus03a}. The case of YbAgGe
appears to have the potential of being somewhat similar to
YbRh$_2$Si$_2$: the relatively high value of $\gamma$ and the
proximity of the magnetic ordering temperature to $T$ = 0 suggest
that YbAgGe is close to a quantum critical point (QCP) and make it
a good candidate for a study of the delicate balance and
competition between magnetically ordered and strongly correlated
ground states under the influence of a number of parameters such
as pressure, chemical substitution and/or magnetic field.

In this work we report the magnetic-field-induced evolution of the
ground state of YbAgGe as seen in anisotropic resistivity and
specific heat measurements up to 140 kOe. We show that on increase
of the applied magnetic field the progression from small moment
magnetic order to QCP with the evidence of non-Fermi-liquid (NFL)
behavior and, in higher fields, to low temperature Fermi-liquid
(FL) state is observed.

\section{Experimental}
YbAgGe crystallizes in hexagonal ZrNiAl-type structure
\cite{gib96a,pot97a}. YbAgGe single crystals in the form of clean
hexagonal cross section rods of several mm length and 0.3-0.8
mm$^2$ cross section were grown from high temperature ternary
solutions rich in Ag and Ge. Their structure and the absence of
impurity phases were confirmed by powder X-ray diffraction (see
\cite{mor03a} for details of the samples' growth). Temperature and
field dependent resistivity $\rho(H,T)$ and heat capacity
$C_p(H,T)$ were measured down to 0.4 K in an applied magnetic
field up to 140 kOe in a Quantum Design PPMS-14 instrument with
He-3 option. For resistivity a standard ac four probe resistance
technique ($f$ = 16 Hz, $I$ = 1-0.3 mA) was used. Pt leads were
attached to the sample with Epotek H20E silver epoxy so that the
current was flowing along the crystallographic $c$ axis. For these
measurements the magnetic field was applied in two directions: in
$(ab)$ plane, approximately along $[1\bar{2}0]$ direction
(transverse, $H \perp I$, magnetoresistance) and along $c$-axis
(longitudinal, $H \parallel I$ magnetoresistance). For heat
capacity measurements a relaxation technique with fitting of the
whole temperature response of the microcalorimeter was utilized.
The background heat capacity (sample platform and grease) was
measured for all necessary $(H,T)$ values and was accounted for in
the final results. $C_p(H,T)$ was also measured with the field
along the $c$-axis and in the $(ab)$ plane (the same orientations
as in $\rho(H,T)$ data). The heat capacity of LuAgGe was measured
in the same temperature range and was used to estimate a
non-magnetic contribution to the heat capacity of YbAgGe.

\section{Results}
\subsection{\label{Hab}$H \parallel (ab)$}
\begin{figure}
\begin{center}
\includegraphics[angle=0,width=80mm]{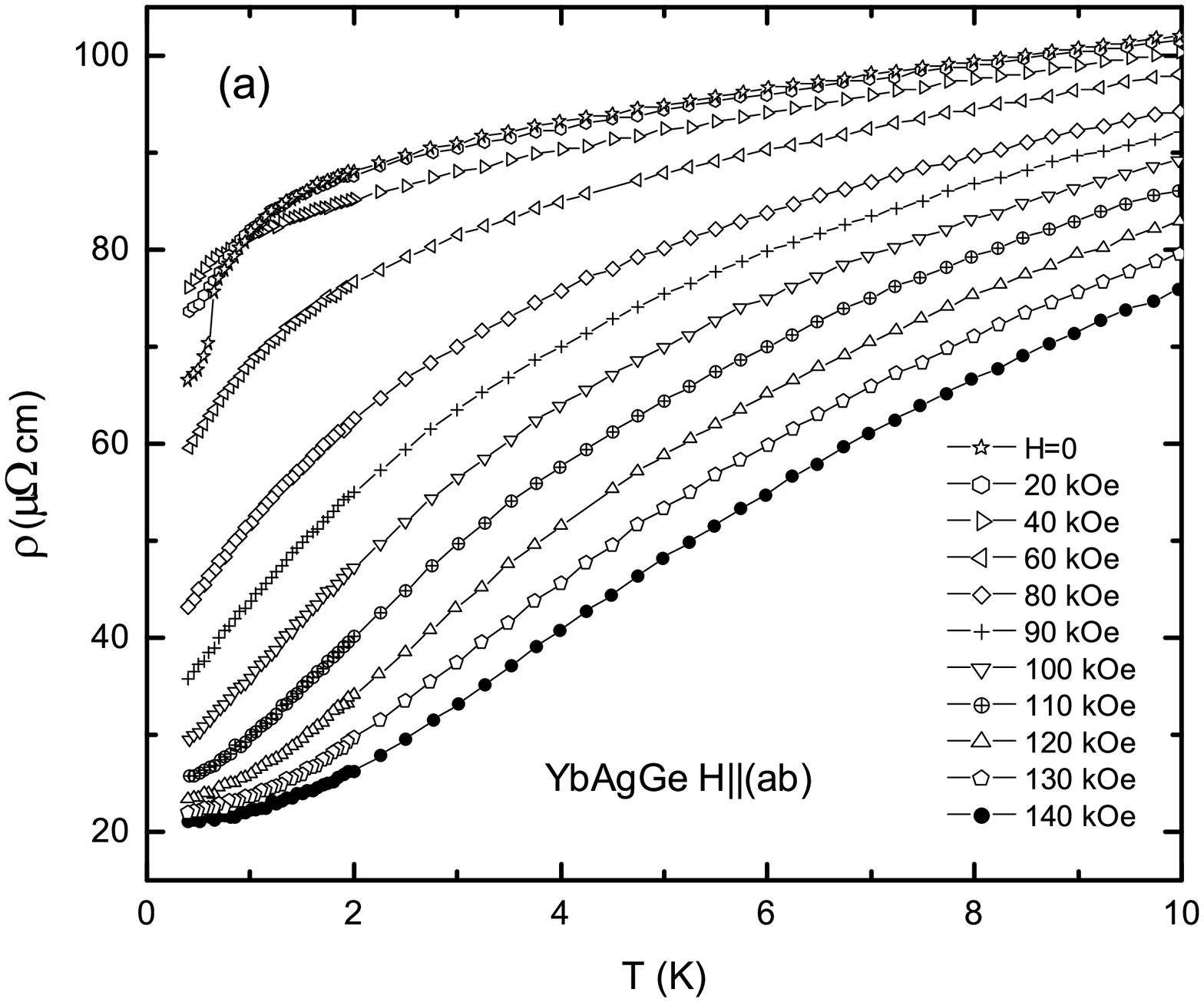}
\includegraphics[angle=0,width=80mm]{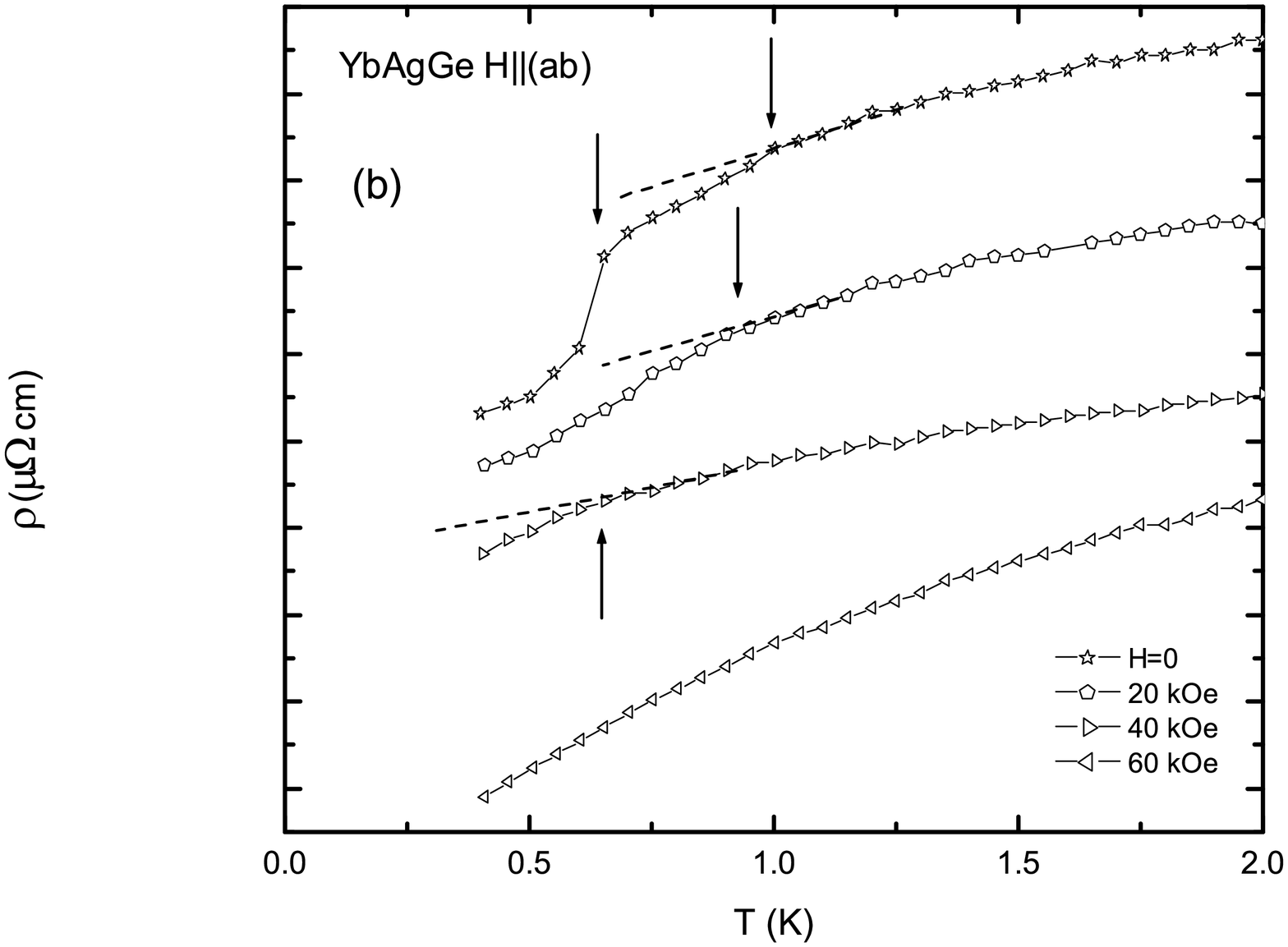}
\includegraphics[angle=0,width=80mm]{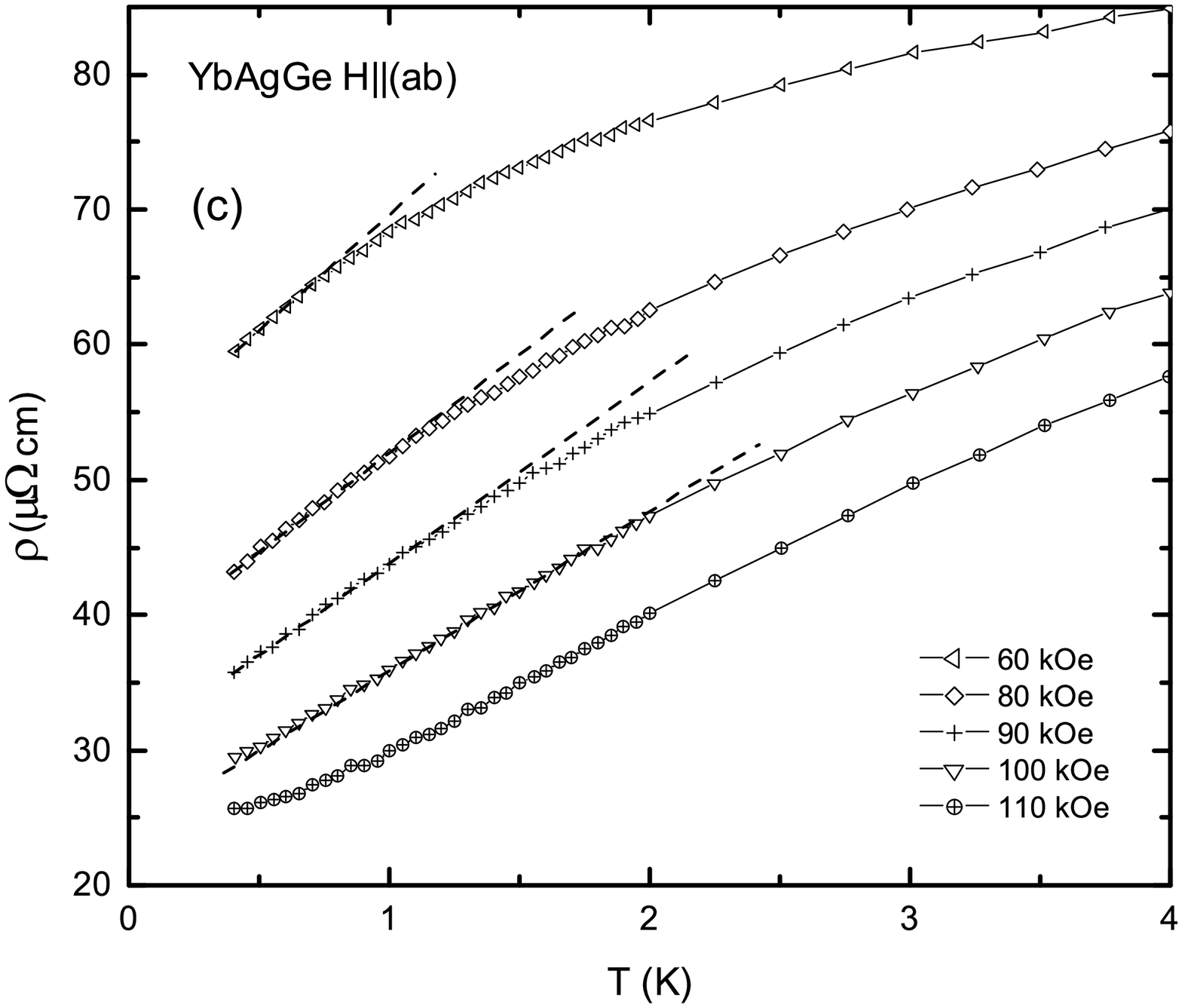}
\includegraphics[angle=0,width=80mm]{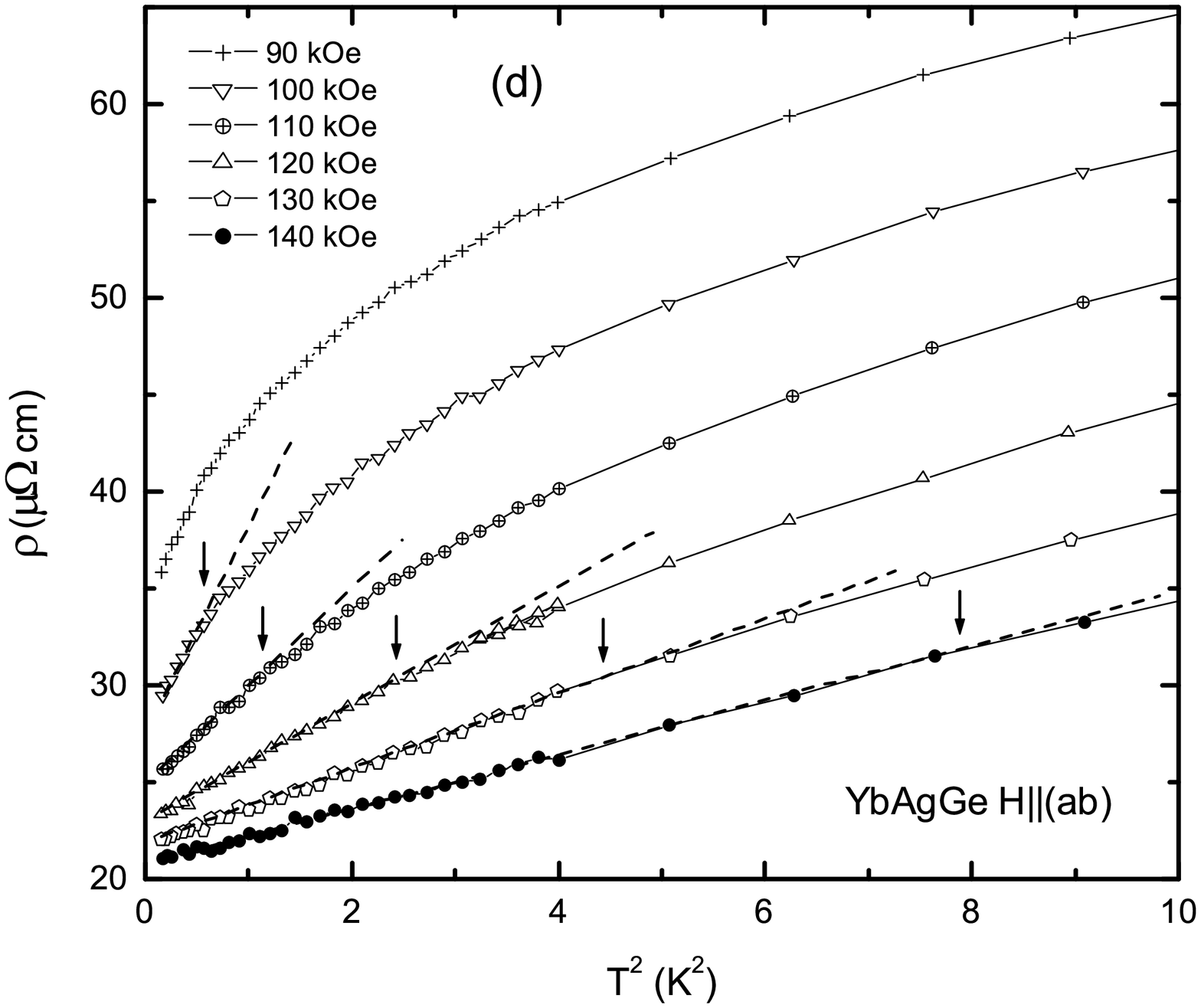}
\end{center}
\caption{(a)Low temperature part of $\rho(T)$ curves for YbAgGe
taken at different applied fields $H \parallel (ab)$; (b)$\rho(T)$
for $H =$ 0, 20, 40 and 60 kOe below 2 K (curves shifted along $y$
axis for clarity), arrows indicate possible magnetic ordering
transitions; (c)$\rho(T)$ for $H =$ 60, 80, 90, 100 and 110 kOe
below 4 K, dashed lines are guides for the eye emphasizing regions
of linear $\rho(T)$; (d)resistivity at $H =$ 90, 100, 110, 120,
130 and 140 kOe below $\sim$ 3 K as a function of $T^2$, dashed
lines bring attention to the regions where $\rho(T) = \rho_0 + A
T^2$, arrows indicate temperatures at which deviations from $T^2$
behavior occur.}\label{RTab}
\end{figure}

The low temperature part of the temperature dependent resistivity
measured in various constant magnetic fields applied in the $(ab)$
plane is shown in Fig. \ref{RTab}(a). There are several features
that apparently require detailed examination. Multiple transitions
in zero field is a feature that is common throughout the RAgGe
series \cite{mor03a} and in the case of YbAgGe in zero field in
addition to the sharp transition at approximately 0.64 K another,
albeit less pronounced feature is apparent at $\sim$ 1 K. Whereas
the 0.64 K transition seems to be pushed below the base
temperature of our measurements by a 20 kOe field, the other
feature shifts down in temperature more gradually and is seen up
to 40 kOe. For applied field between 60 and 90 kOe (Fig.
\ref{RTab}(c)) low temperature $\rho(T)$ functional dependence is
linear down to our base temperature, with the upward curvature in
$\rho(T)$ starting to occur below $\sim 0.8$ K in 100 kOe field.
In higher applied fields ($H \geq 100 kOe$) (Fig. \ref{RTab}(d))
low temperature resistivity follows $\rho(T)= \rho_0 + AT^2$
(Fermi-liquid-like) functional behavior with the range of its
occurrence (for each curve the upper limit is marked with arrow in
Fig. \ref{RTab}(d)) increasing with the increase of applied field
and the coefficient $A$ decreasing.

\begin{figure}
\begin{center}
\includegraphics[angle=0,width=100mm]{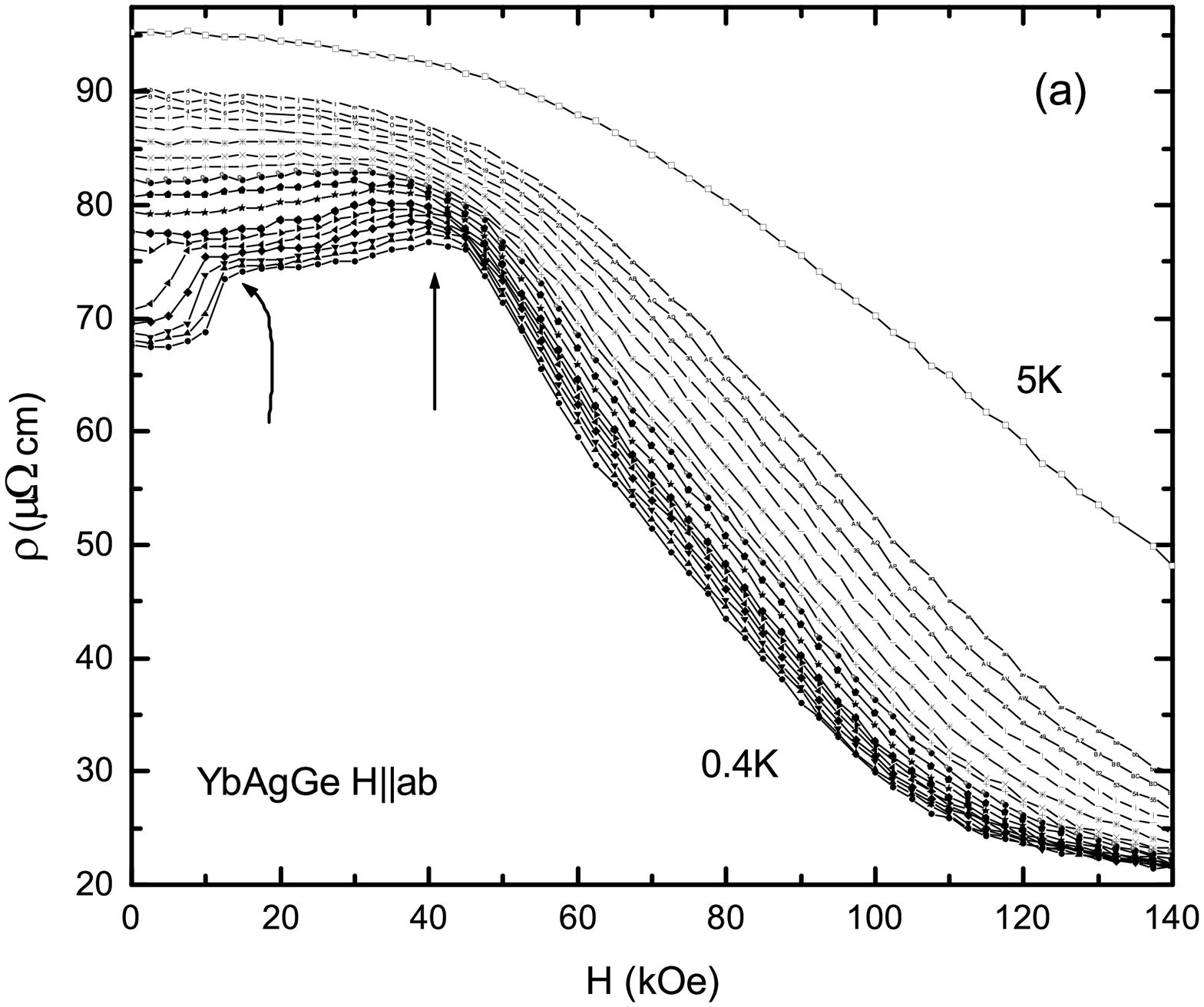}
\includegraphics[angle=0,width=100mm]{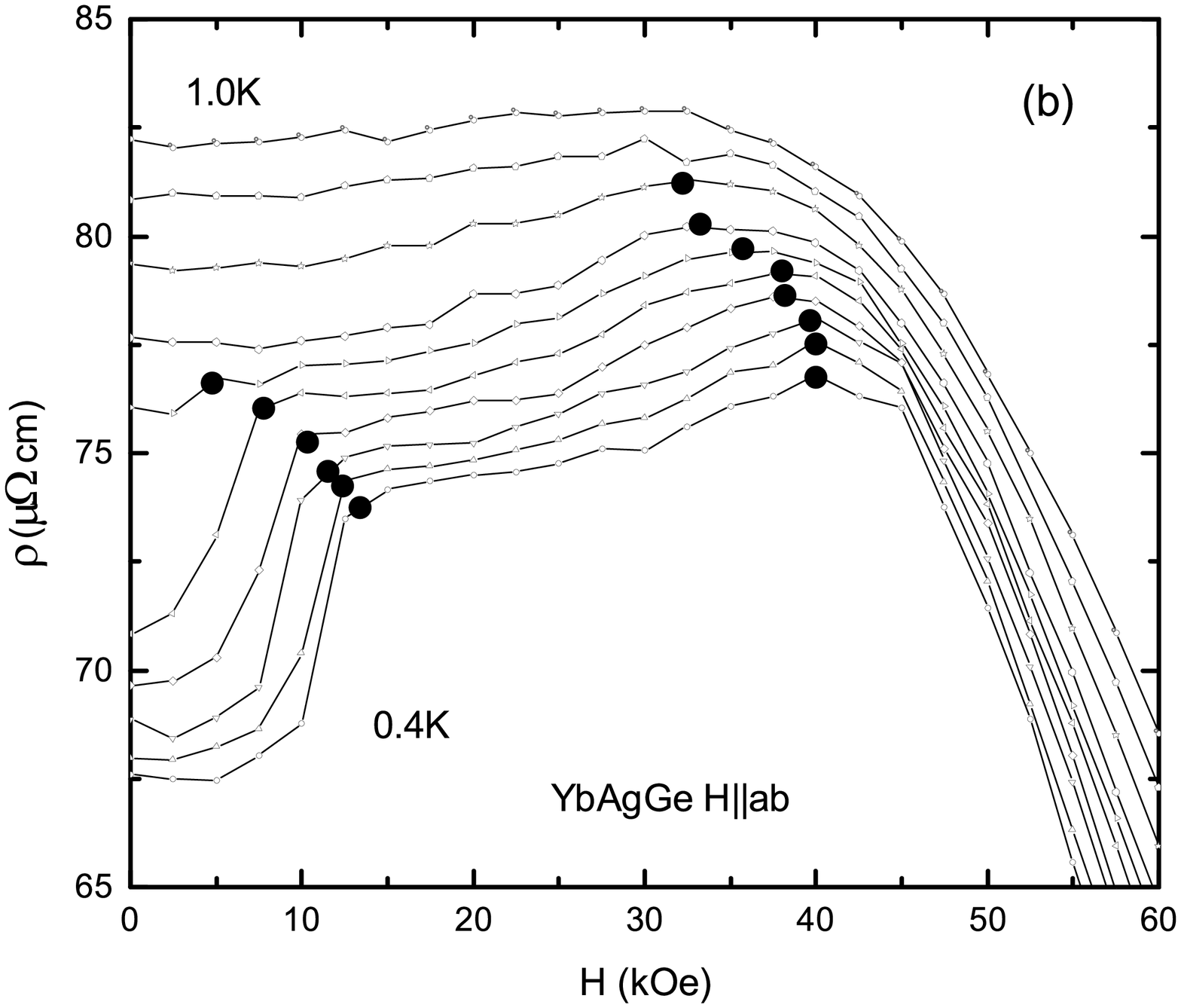}
\end{center}
\caption{(a)$\rho(H)$ ($H \parallel (ab)$) isotherms for YbAgGe
taken every 0.05 K between 0.4 K and 0.7 K, every 0.1 K between
0.7 K and 1.2 K, every 0.2 K between 1.2 K and 2.0 K and at 2.3 K,
2.5 K and 5.0 K, arrows point to the transitions discussed in the
text; (b)enlarged low field - low temperature (0-60 kOe, 0.4-1.0
K) part of the panel (a), black dots mark transitions on the
respective curves.}\label{RHab}
\end{figure}

Field dependent resistivity data taken at constant temperatures
between 0.4 K and 5.0 K are shown in Fig. \ref{RHab}(a). At $T$ =
0.4 K two fairly sharp features, at $\sim$ 13 kOe and $\sim$ 40
kOe are seen in the $\rho(H)$ data. The lower field feature may be
identified as a signature of a metamagnetic transition between two
different magnetically ordered phases. This feature vanishes as
the temperature increased to $T >$ 0.65 K (Fig. \ref{RHab}(b)).
The second, more smoothed, higher field feature, may be to a
transition from a magnetically ordered state to a saturated
paramagnetic state. If the critical field for this transition is
inferred from the maximum in $\rho(H)$, this transition can be
discerned up to 0.8-0.9 K (Fig. \ref{RHab}(b)), consistent with it
being associated with the $\sim 1$ K transition seen in the $H =
0$ $\rho(T)$ and $C_p(T)$ data (Fig. \ref{RTCT}). At higher
temperatures (Fig. \ref{RHab}(a)) this feature broadens and
resembles a crossover rather than a transition. For 2 K $< T <$ 5
K $\rho(H)$ looks like a generic magnetoresistance of a
paramagnetic metal \cite{yos57a}. The large black dots in Fig.
\ref{RHab}(b) show the evolution of the two aforementioned
features.

\begin{figure}
\begin{center}
\includegraphics[angle=0,width=80mm]{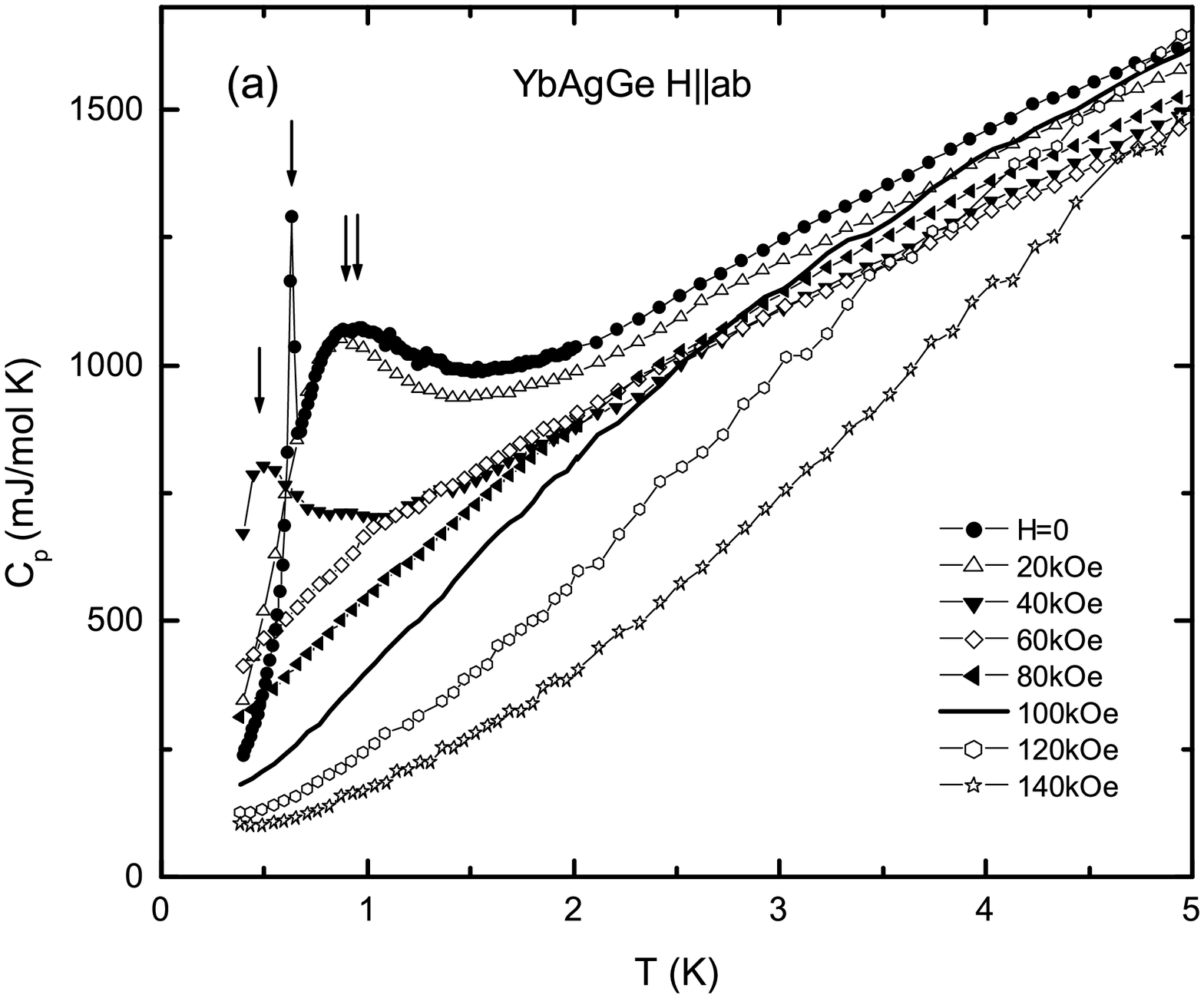}
\includegraphics[angle=0,width=80mm]{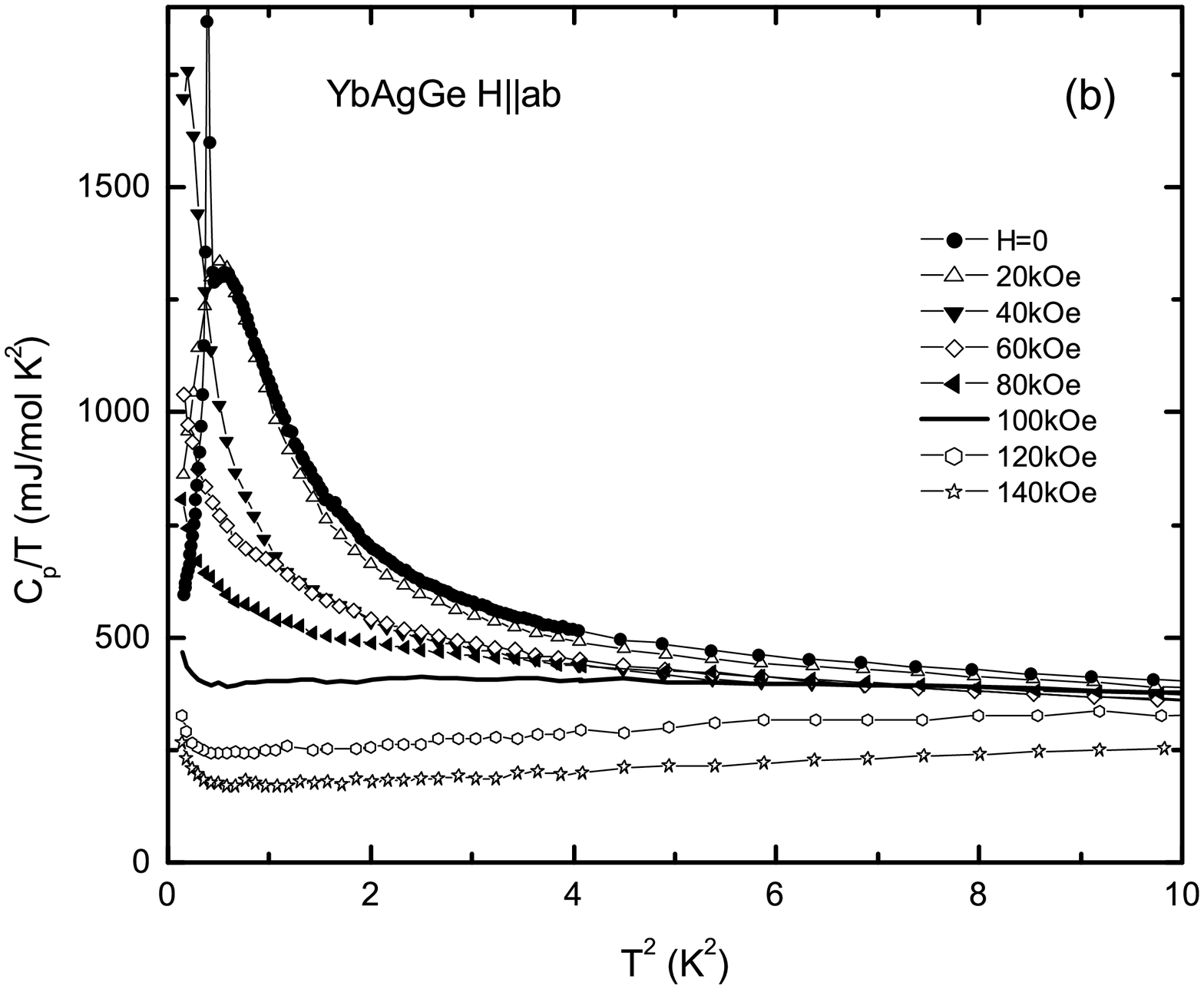}
\includegraphics[angle=0,width=80mm]{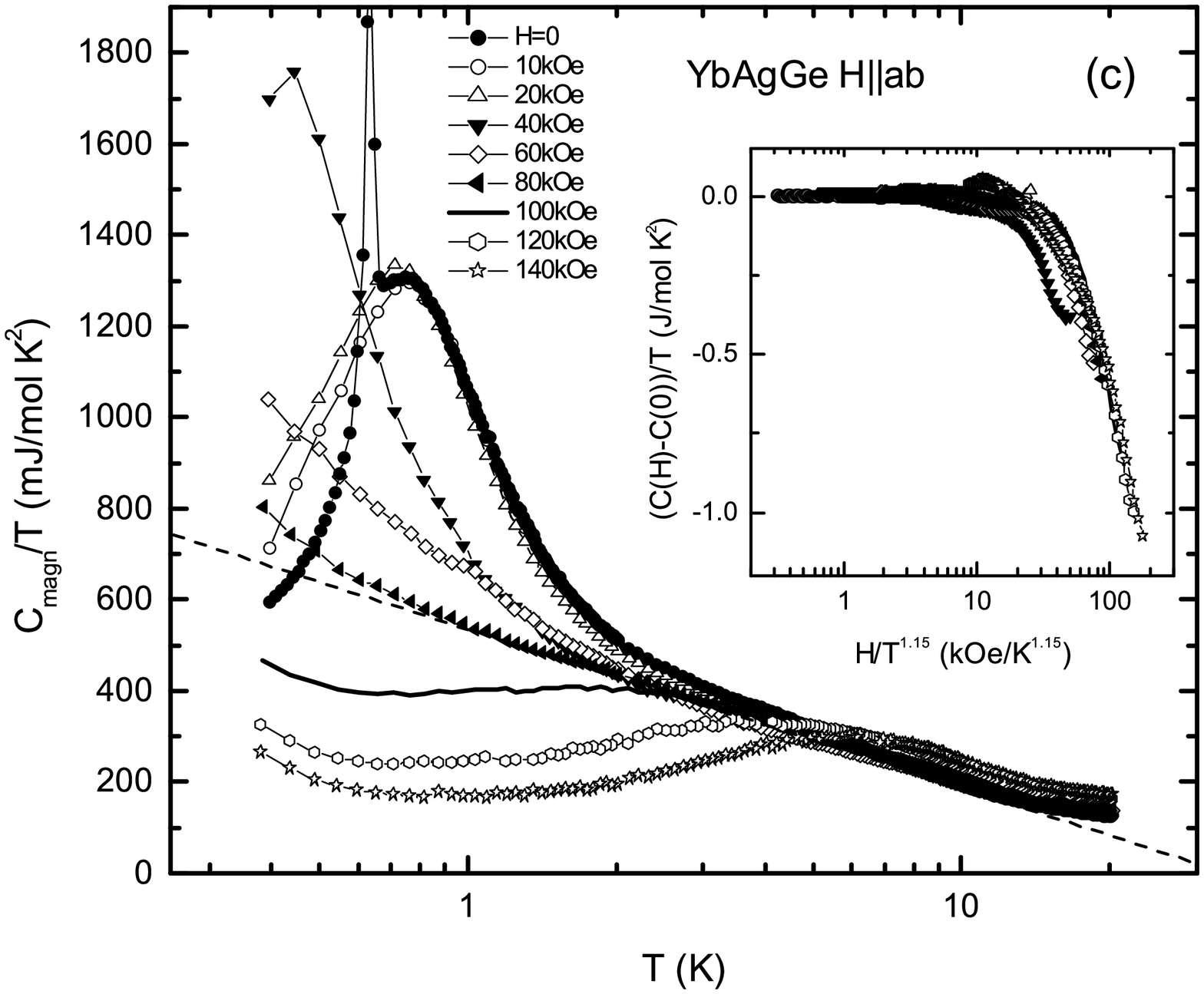}
\end{center}
\caption{(a)Low temperature part of the heat capacity curves for
YbAgGe taken at different applied fields $H \parallel (ab)$,
arrows indicate peaks associated with magnetic ordering; (b)low
temperature part of $C_p$ {\it vs} $T^2$ curves; (c){\it semi-log}
plot of the magnetic part ($C_{magn} = C_p(YbAgGe)-C_p(LuAgGe)$)
of the heat capacity, $C_{magn}/T$ {\it vs} $T$, for different
applied magnetic fields, dashed line is a guide to the eye, it
delineates linear region of the $H$ = 80 kOe curve; inset: {\it
semi-log} plot of $(C(H)-C(H=0))/T$ {\it vs} $H/T^{1.15}$ ($T
\geq$ 0.8 K), note approximate scaling of the data for $H \geq$ 60
kOe.}\label{HCab}
\end{figure}

The low temperature heat capacity of YbAgGe is shown for several
values of applied magnetic field in Fig. \ref{HCab}(a). The lower
temperature, sharp peak is seen only for $H =$ 0, having been
suppressed below the base temperature by an applied field of 20
kG. The higher temperature maximum seen just below 1 K (for $H =$
0) shifts down with the increase of the applied field and drops
below 0.4 K for $H \geq 60 kOe$. The field dependence of this
feature is consistent with that of the higher temperature feature
in resistivity discussed above giving further evidence that YbAgGe
has two closely-spaced magnetic transitions. The same data plotted
as $C_p/T$ {\it vs} $T^2$ (Fig. \ref{HCab}(b)) allow for the
tracking of the variation of the electronic specific heat
coefficient $\gamma$ in applied field (for $H \geq$ 60 kOe, when
the magnetic order is suppressed). The values at $T^2$ = 0.35
K$^2$ (a value chosen to avoid the upturn in lowest temperature,
highest field $C_p/T(T^2)$ data possibly associated with the
nuclear Schottky contribution) give a reasonable approximation of
$\gamma(H)$. A more than four-fold decrease of $\gamma$ is
observed from 60 kOe to 140 kOe.

The magnetic contribution to the YbAgGe specific specific heat
(defined as $C_{magn} = C_p(YbAgGe) - C_p(LuAgGe)$) is shown in
Fig. \ref{HCab}(c) in $C_{magn}/T$ {\it vs} $\lg T$ coordinates.
(It should be noted that $C_p(T)$ of LuAgGe was measured at $H$ =
0 and 140 kOe and found to be insensitive to the applied field in
this temperature range.) For intermediate values of applied field
there is a region of the logarithmic divergency seen in the
specific heat data $C_{magn}/T \propto -\ln T$. The largest range
of the logarithmic behavior (more than an order of magnitude in
temperature, from below 1 K to above 10 K) is observed for $H = $
80 kOe. These data can be described as $C_{magn}/T = \gamma
\prime_0 \ln (T_0/T)$ with $\gamma \prime_0 \approx$ 144 mJ/mol
K$^2$ and $T_0 \approx$ 41 K. These parameters are of the same
order of magnitude as those reported for YbRh$_2$Si$_2$
\cite{tro00a}. In higher fields Fermi-liquid-like behavior
apparently recovers, in agreement with the resistivity data.

The crossover function $(C(H)-C(H=0))/T$ {\it vs} $H/T^\beta$
($\beta$ = 1.15) (one of the expressions considered in the scaling
analysis at a QCP) is shown in the inset to Fig. \ref{HCab}(c).
Data for $H \geq$ 60 kOe collapse onto one universal curve. Such
scaling behavior \cite{tsv93a} with $\beta$ between 1.05 and 1.6
was observed for a number of materials that demonstrate NFL
properties \cite{tro00a,and91a,len97a,heu98a,koe00a} and may be
considered as further corroboration of the proximity of YbAgGe to
a QCP.

\subsection{\label{Hc}$H \parallel c$}
\begin{figure}
\begin{center}
\includegraphics[angle=0,width=80mm]{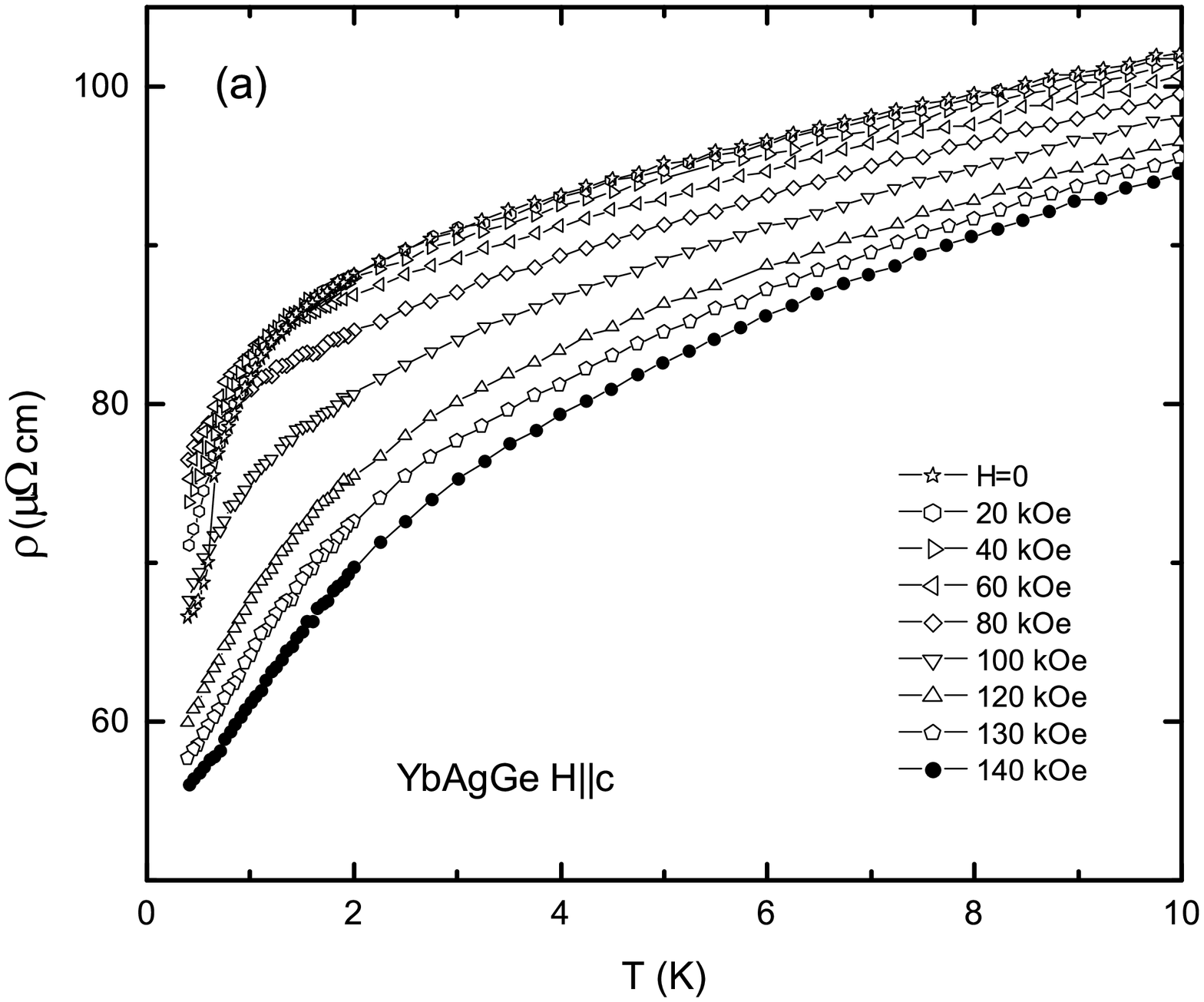}
\includegraphics[angle=0,width=80mm]{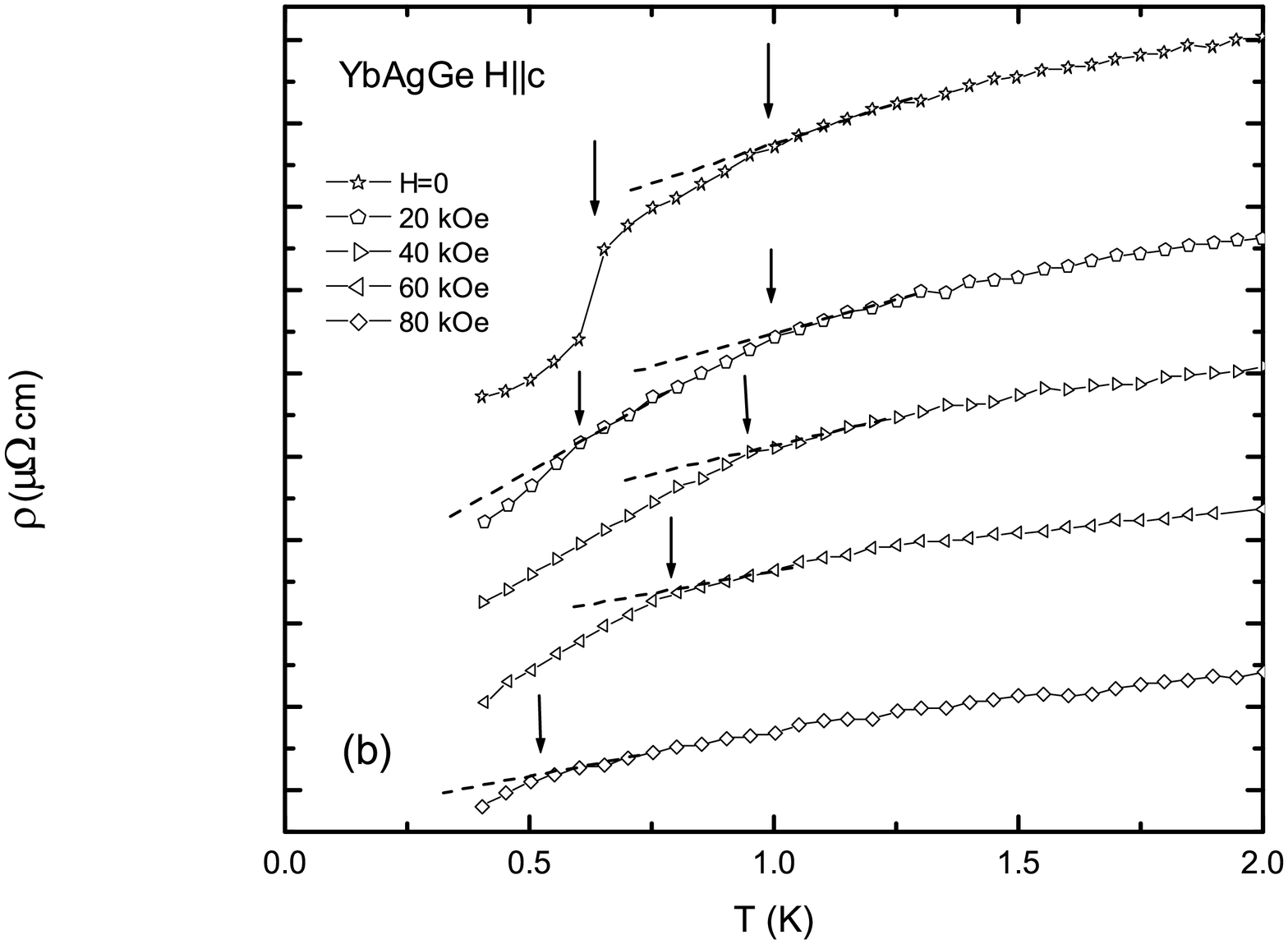}
\includegraphics[angle=0,width=80mm]{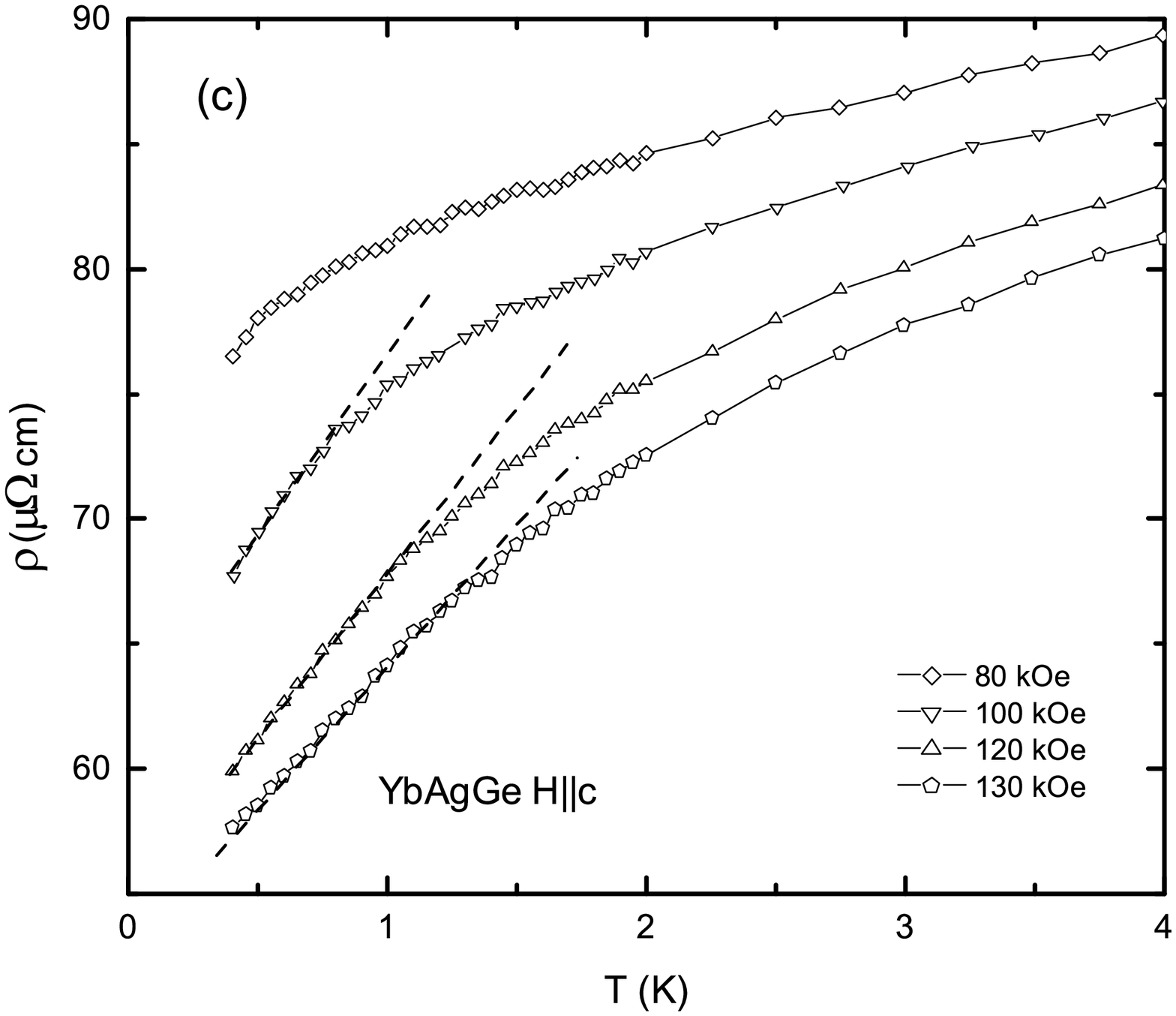}
\includegraphics[angle=0,width=80mm]{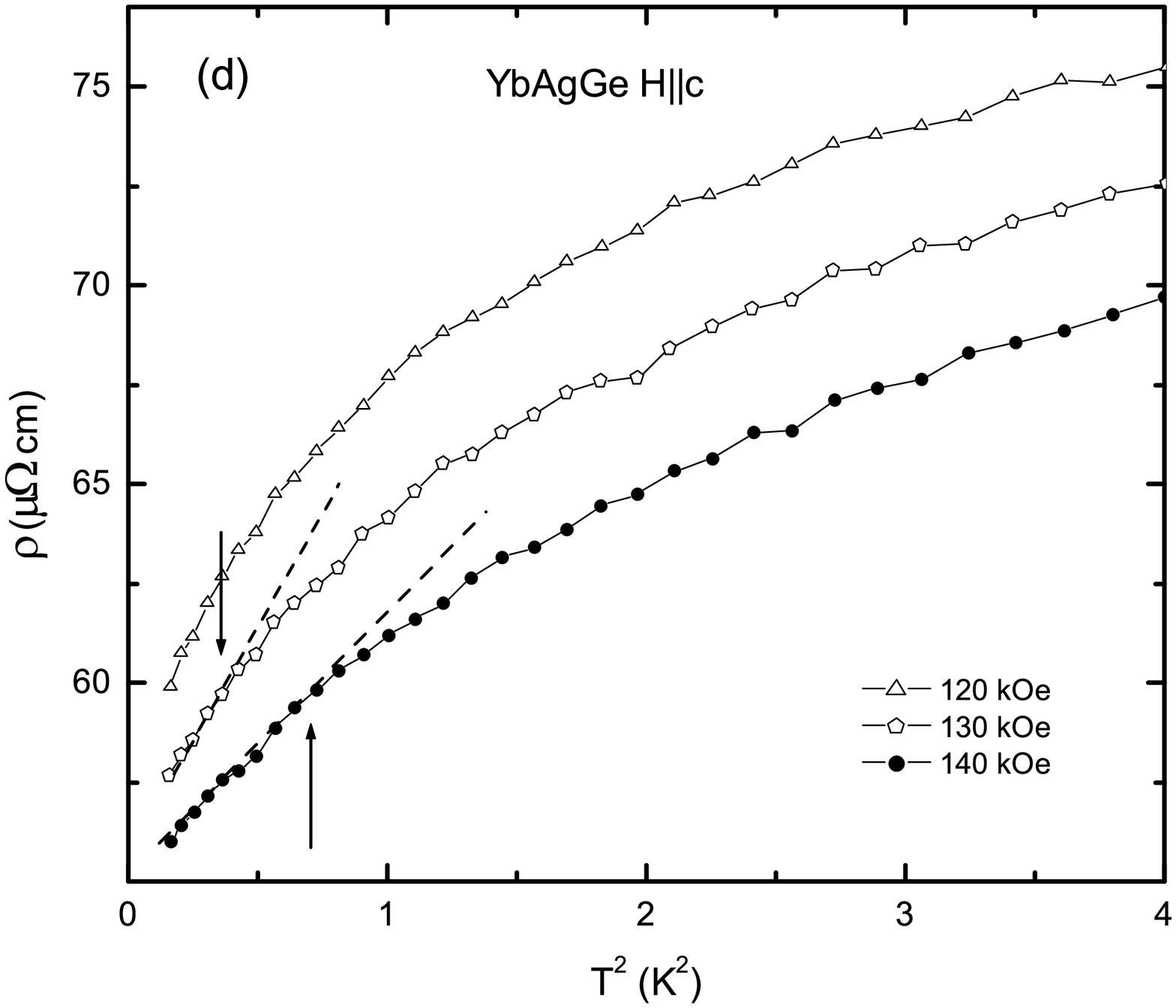}
\end{center}
\caption{(a)Low temperature part of $\rho(T)$ curves for YbAgGe
taken at different applied fields $H \parallel c$; (b)$\rho(T)$
for $H = 0-80$ kOe below 2 K (curves shifted along $y$ axis for
clarity), arrows indicate magnetic ordering transitions;
(c)$\rho(T)$ for $H =$ 80, 100, 120 and 130 kOe below 4 K, dashed
lines are guides for the eye emphasizing regions of linear
$\rho(T)$; (d)resistivity at $H =$  120, 130 and 140 kOe below
$\sim$ 3 K as a function of $T^2$, dashed lines bring attention to
the regions where $\rho(T) = \rho_0 + A T^2$, arrows indicate
temperatures at which deviations from $T^2$ behavior
occur.}\label{RTc}
\end{figure}

\begin{figure}
\begin{center}
\includegraphics[angle=0,width=100mm]{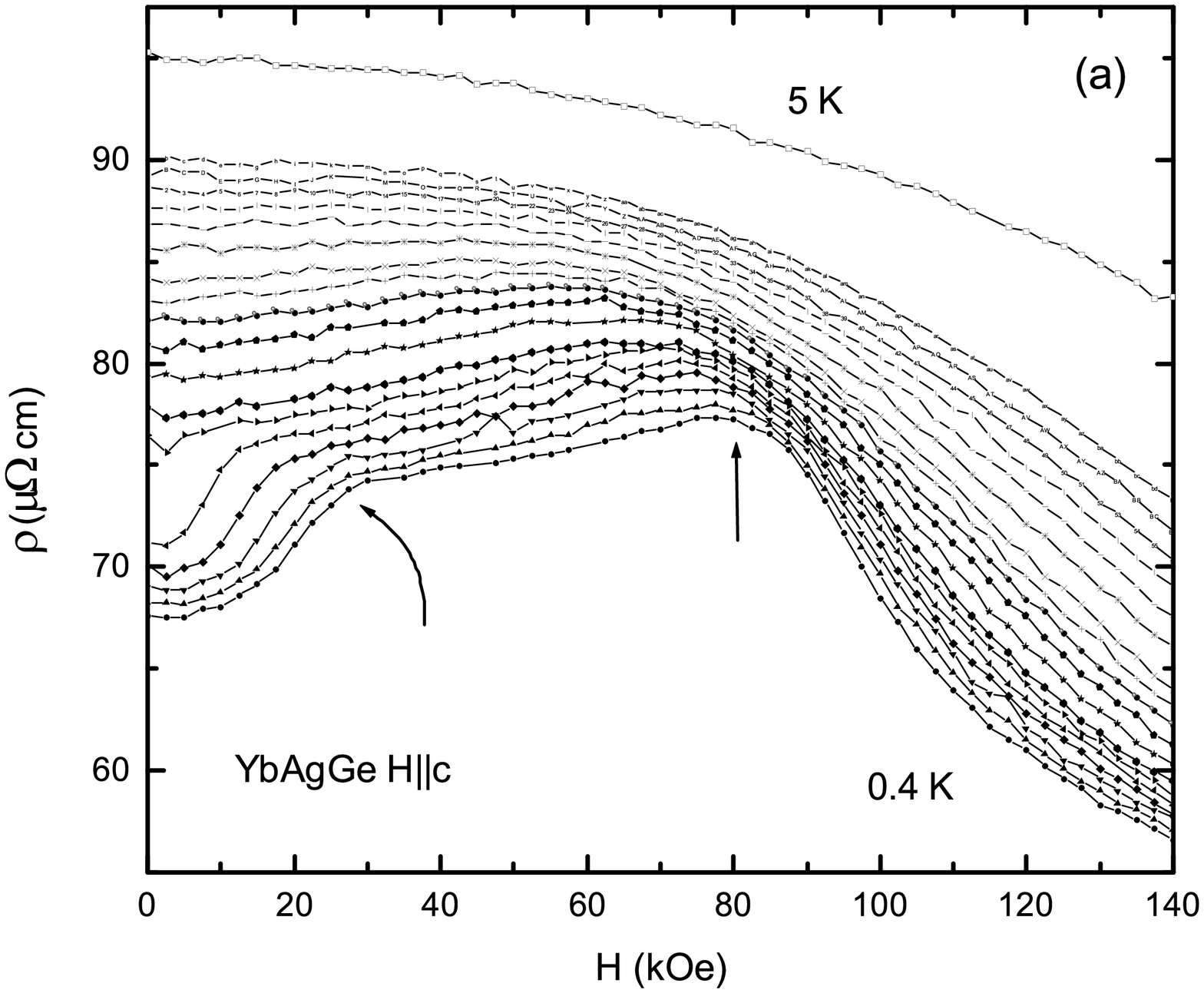}
\includegraphics[angle=0,width=100mm]{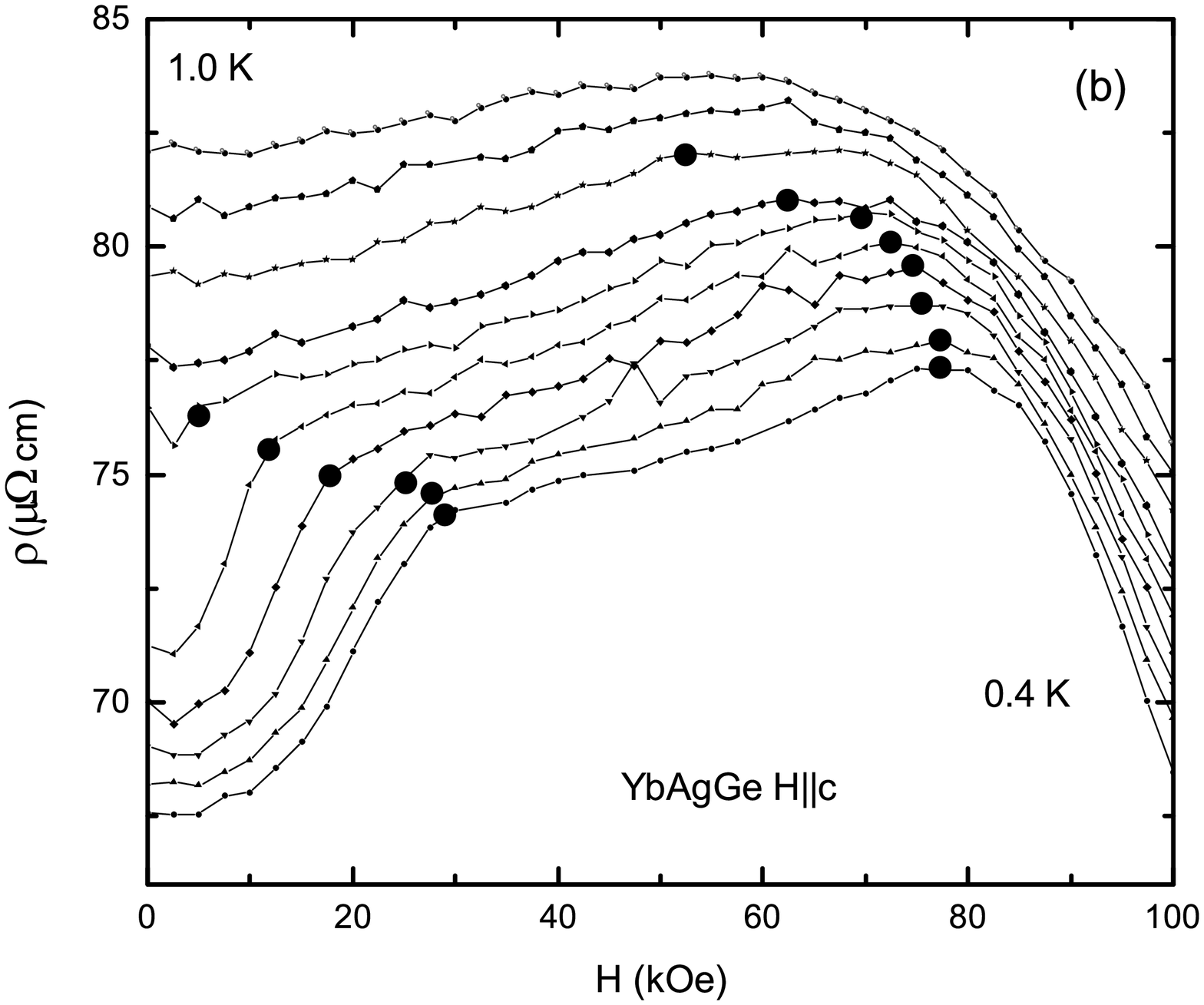}
\end{center}
\caption{(a)$\rho(H)$ ($H \parallel c$) isotherms for YbAgGe taken
every 0.05 K between 0.4 K and 0.7 K, every 0.1 K between 0.8 K
and 1.2 K, every 0.2 K between 1.4 K and 2.0 K and at 2.3 K, 2.5 K
and 5.0 K, arrows point to the transitions discussed in the text;
(b)enlarged low field - low temperature (0-100 kOe, 0.4-1.0 K)
part of the panel (a), black dots mark transitions on the
respective curves.}\label{RHc}
\end{figure}

YbAgGe manifests an easy plane anisotropy in both its low-field
and high-field magnetization (Fig. \ref{MTH}). This is a trend
that evolves across the RAgGe series \cite{mor03a}, {\it e.g.} in
TmAgGe the local moments are extremely anisotropic being confined
to the basal plane. Not surprisingly this anisotropy manifests
itself in the low-temperature $\rho(H,T)$ and $C_p(H,T)$ data. The
variation of the temperature dependent resistivity for $H
\parallel c$ (Fig. \ref{RTc}(a)) is comparable to that for $H
\parallel (ab)$. As for the in-plane orientation of the field, the two
transitions seen for $H =$ 0 move to lower temperatures with
application of magnetic field, albeit the effect of field is
weaker, so that the lower temperature transition is still being
detected as a break in slope for $H =$ 20 kOe whereas the higher
temperature transition persists up to 80 kOe (Fig. \ref{RTc}(b)).
The field range for which linear, low temperature resistivity can
be seen is smaller and is shifted to higher fields (Fig.
\ref{RTc}(c)), whereas the $\rho - \rho_0 \propto T$ behavior can
be recognized for $H =$ 100 kOe and 120 kOe, and a slight upward
curvature above 0.4 K is already seen at $H =$ 130 kOe.  The low
temperature resistivity can be characterized by $\rho(T) = \rho_0
+ AT^2$ (Fig. \ref{RTc}(d)) and this curvature can be viewed as a
signature of a FL behavior. The range of $T^2$ behavior increases
and the value of $A$ decreases with an increase of applied field.
The field-dependent resistivity for this orientation of the
magnetic field (Fig. \ref{RHc}(a)) is similar to the set of
$\rho(H)$ isotherms for $H \parallel (ab)$ except for the weaker
field dependence of the observed transitions (Fig. \ref{RHc}(b)).

\begin{figure}
\begin{center}
\includegraphics[angle=0,width=80mm]{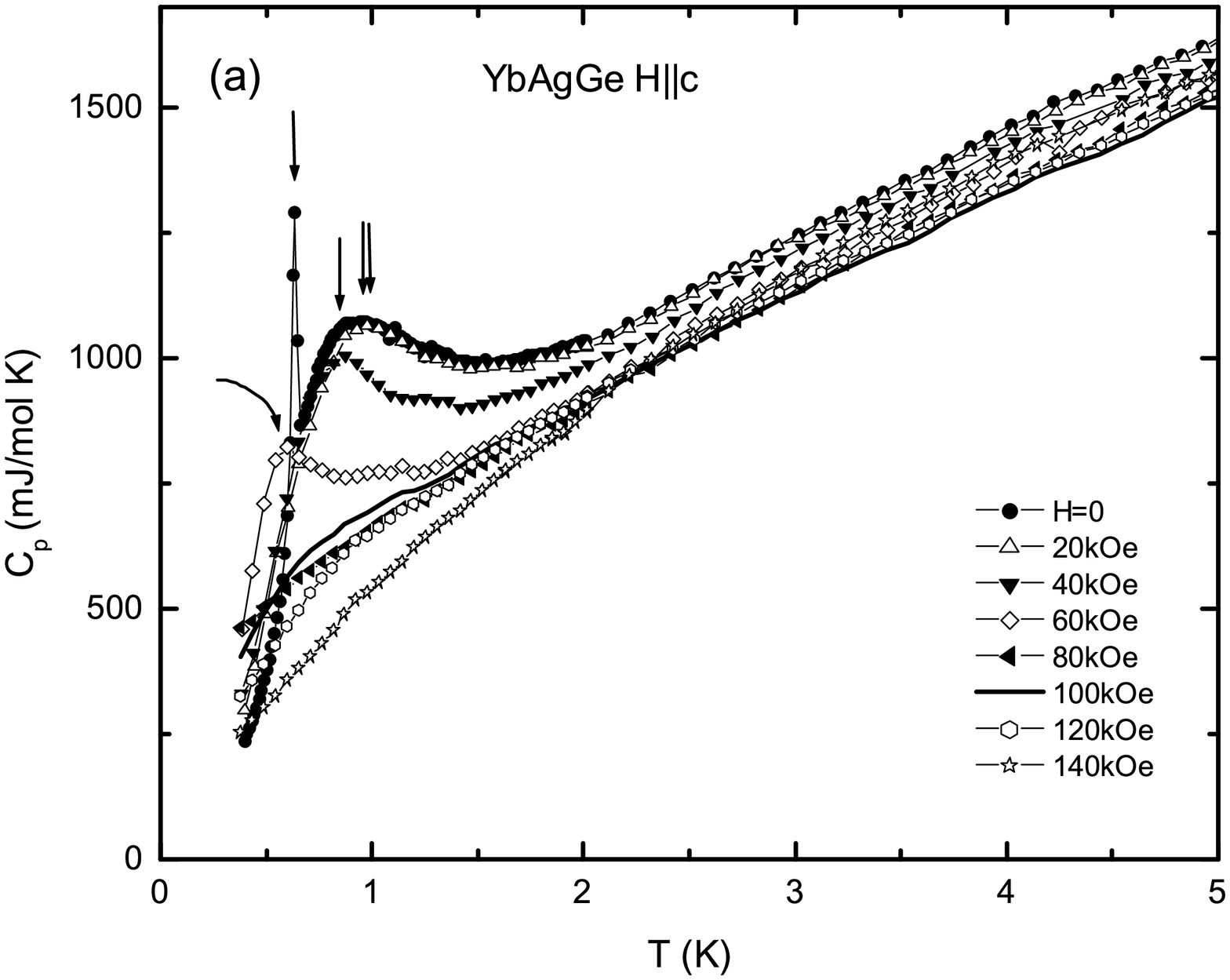}
\includegraphics[angle=0,width=80mm]{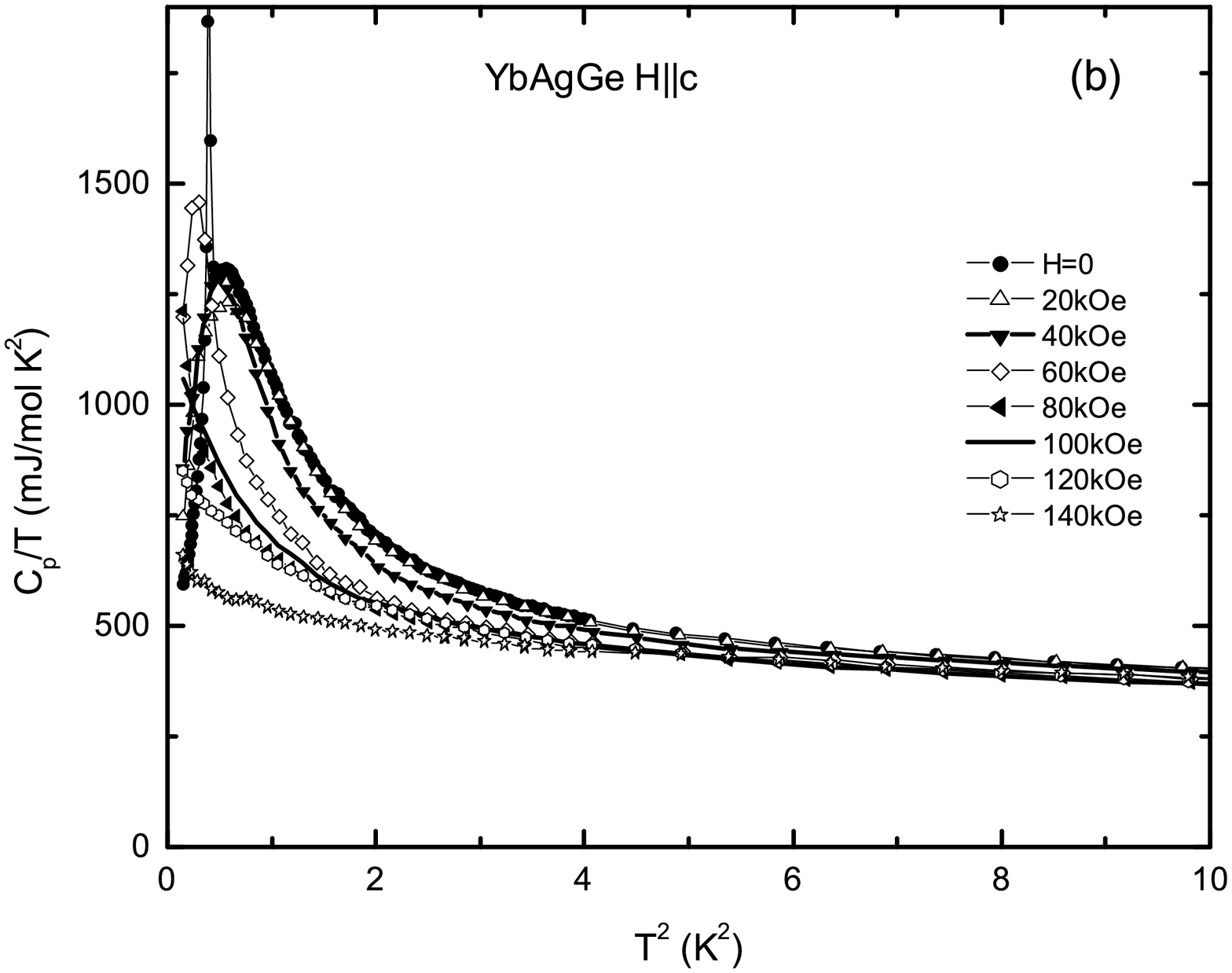}
\includegraphics[angle=0,width=80mm]{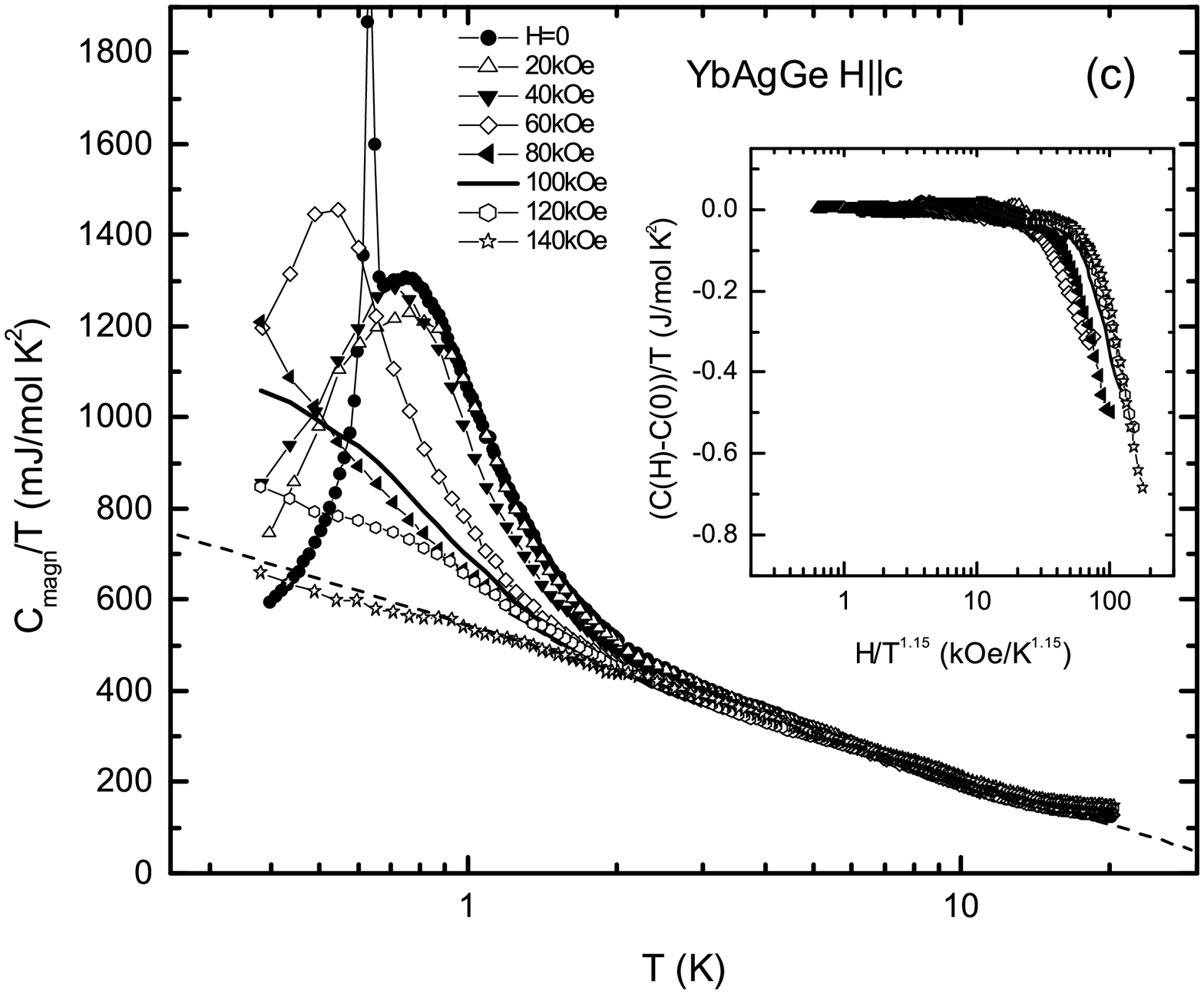}
\end{center}
\caption{(a)Low temperature part of the heat capacity curves for
YbAgGe taken at different applied fields $H \parallel c$, arrows
indicate peaks associated with magnetic ordering; (b)low
temperature part of $C_p$ {\it vs} $T^2$ curves; (c){\it semi-log}
plot of the magnetic part ($C_{magn} = C_p(YbAgGe)-C_p(LuAgGe)$)
of the heat capacity, $C_{magn}/T$ {\it vs} $T$, for different
applied magnetic fields, dashed line is a guide to the eye, it
delineates linear region of the $H$ = 140 kOe curve; inset: {\it
semi-log} plot of $(C(H)-C(H=0))/T$ {\it vs} $H/T^{1.15}$ ($T
\geq$ 0.8 K), note approximate scaling of the data for $H \geq$
100 kOe.}\label{HCc}
\end{figure}

The low temperature part of the heat capacity measured for $H
\parallel c$ up to 140 kOe is shown in Fig. \ref{HCc}(a).
Upper magnetic ordering transition temperature decreases with
increase of applied field and can be followed up to 60 kOe.
Electronic contribution to the specific heat for fields where the
ordering transition is suppressed can be estimated from the Fig.
\ref{HCc}(b). For this orientation the largest range of the
logarithmic behavior $C_{magn}/T \propto -\ln T$ is observed for
$H = $ 140 kOe (Fig. \ref{HCc}(c)) and these data can be expressed
as $C_{magn}/T = \gamma \prime_0 \ln (T_0/T)$ with $\gamma
\prime_0 \approx$ 143 mJ/mol K$^2$ and $T_0 \approx$ 44 K, the
values of $\gamma \prime_0$ and $T_0$ being, within the accuracy
of the data and the fit, the same as for $H \parallel (ab)$.
Scaling behavior of the specific heat data plotted as
$(C(H)-C(H=0))/T$ {\it vs} $H/T^\beta$ (the value of the exponent
$\beta$ = 1.15 is the same as for $H \parallel (ab)$) is observed
for $H \geq$ 100 kOe (Fig. \ref{HCc}(c), inset).

\section{Summary and Discussion}
\begin{figure}
\begin{center}
\includegraphics[angle=0,width=80mm]{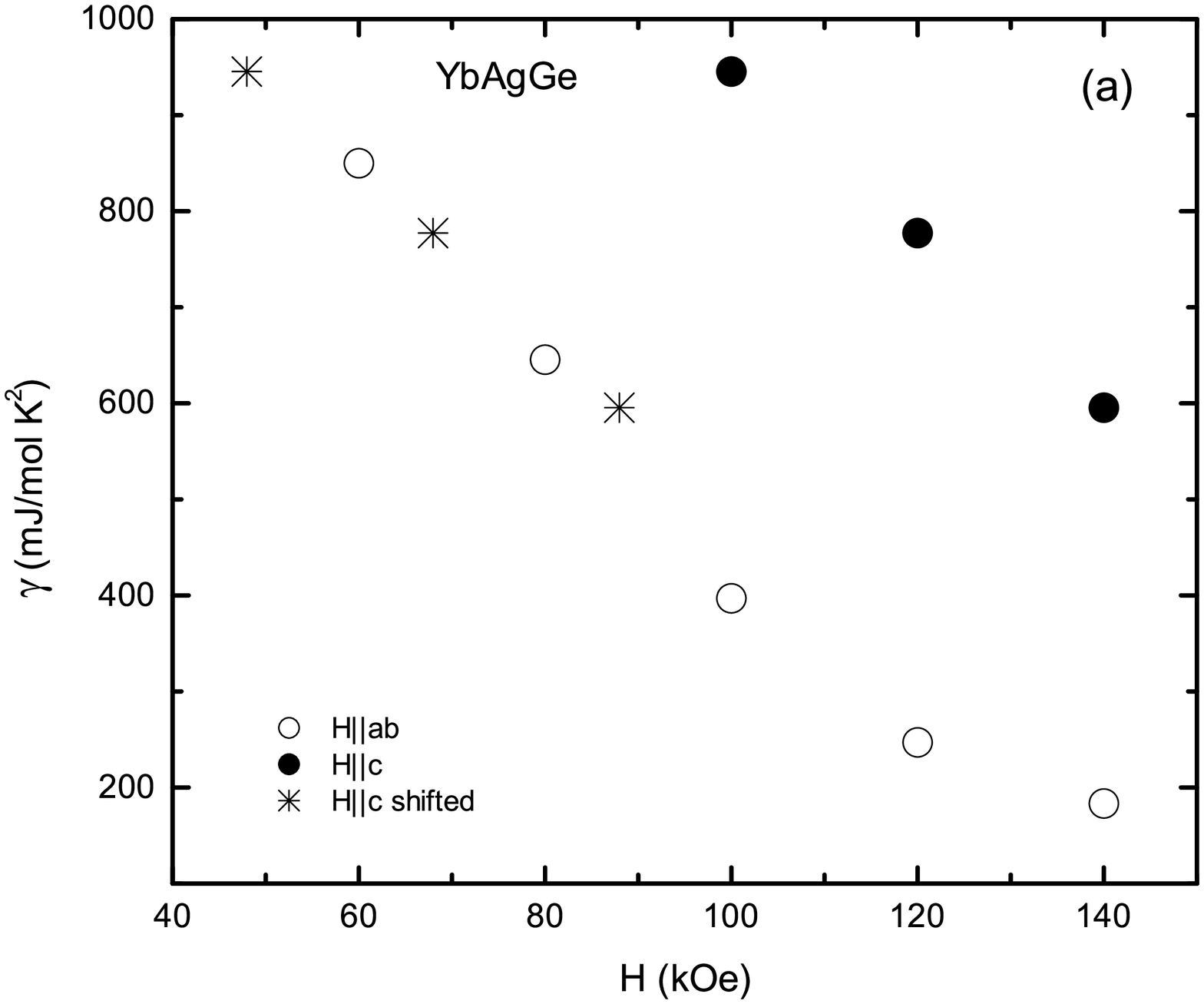}
\includegraphics[angle=0,width=80mm]{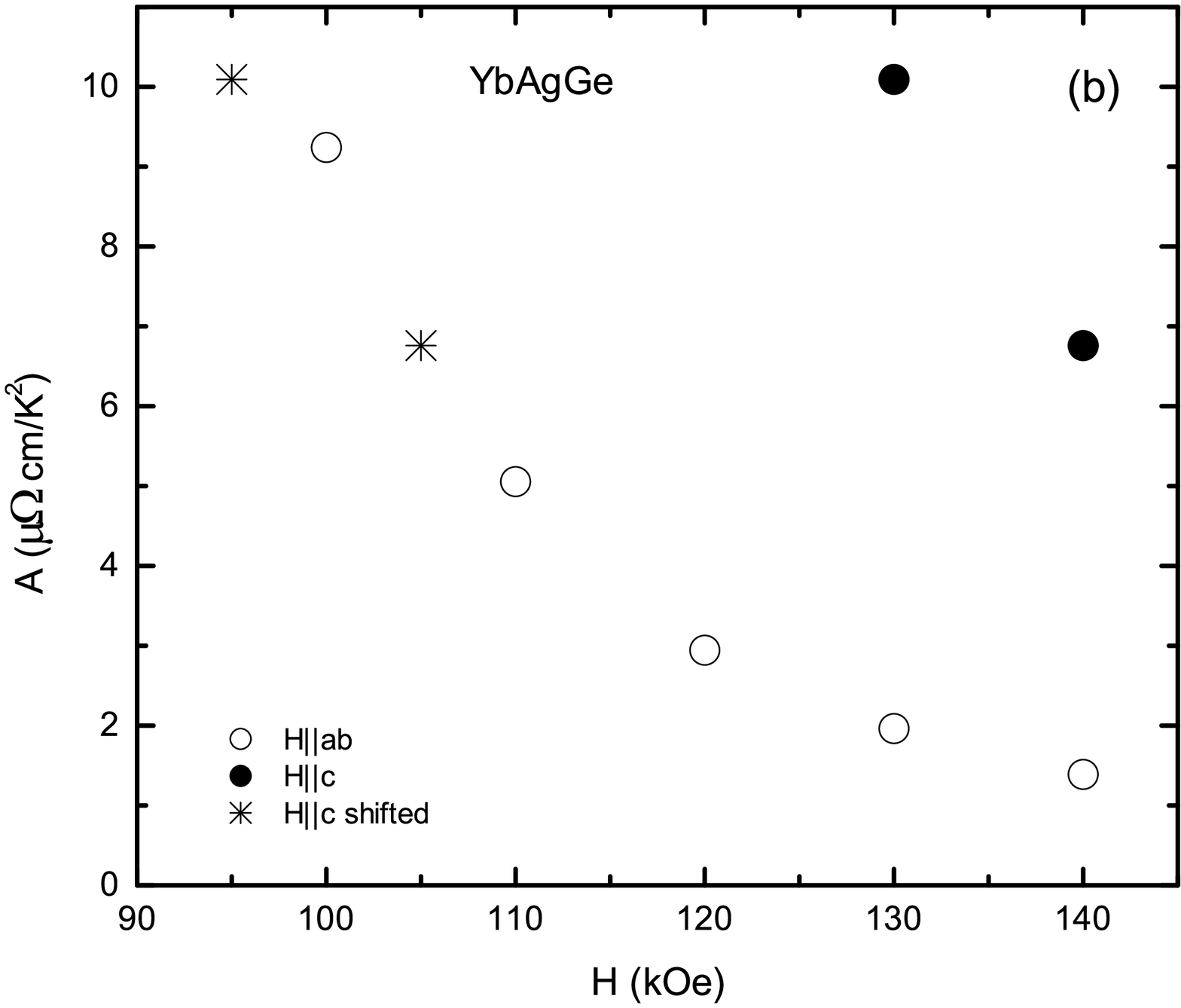}
\includegraphics[angle=0,width=80mm]{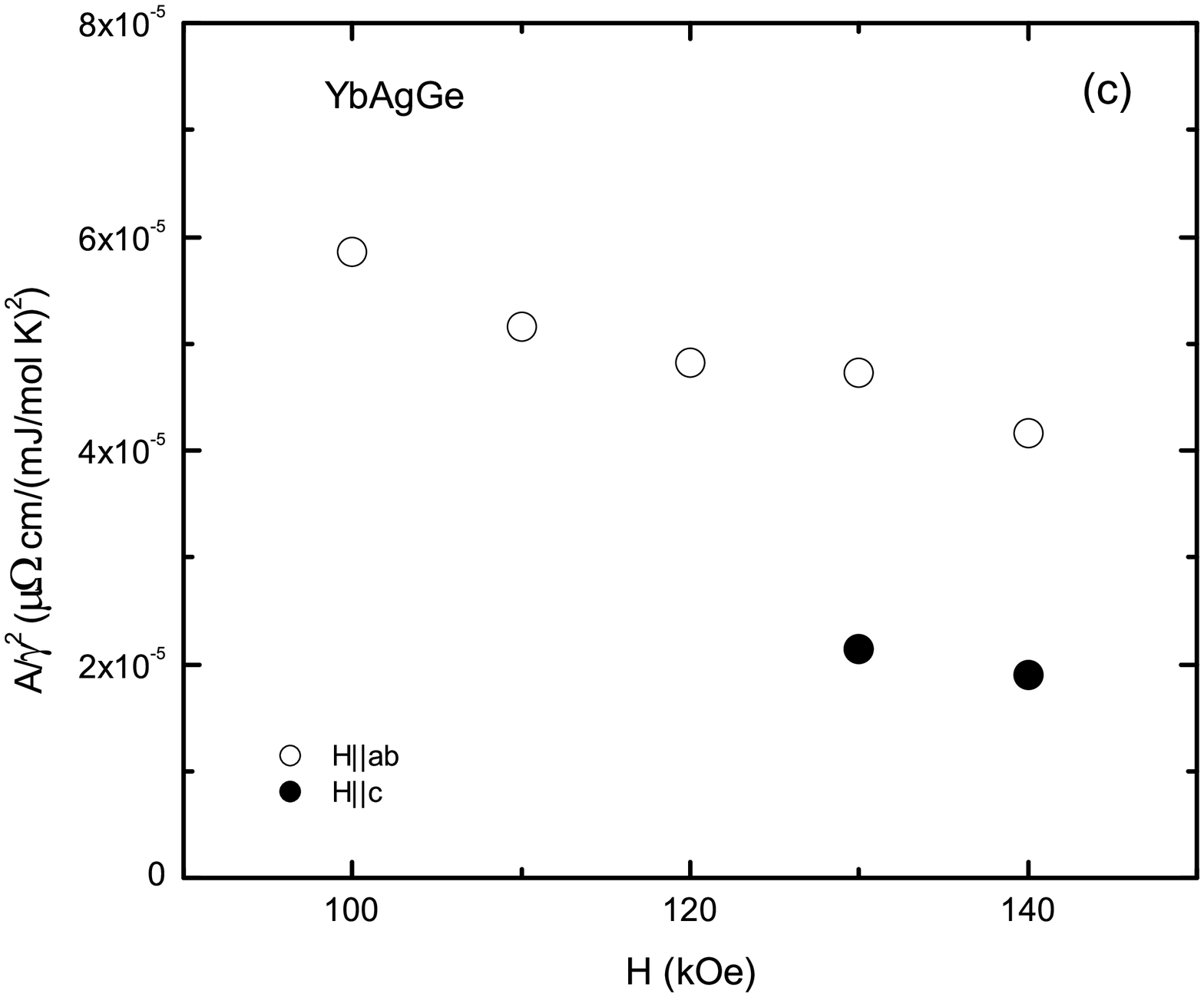}
\end{center}
\caption{The anisotropic field dependence of (a)the electronic
specific heat coefficient $\gamma$; (b)the resistivity coefficient
$A$  and (c)the Kadowaki-Woods ratio $A/\gamma^2$. In figures (a)
and (b) shifted data for $H \parallel c$ are shown as $\ast$ (see
text). The shift for figure (a) is -52 kOe and for figure (b) -35
kOe.}\label{gammaA}
\end{figure}

For both sets of data ($H \parallel (ab)$ and $H \parallel c$), at
high enough applied fields, long range magnetic order is
suppressed, and the electronic contribution to the specific heat
can be estimated, whereas the low temperature resistivity shows
$\Delta \rho \propto AT^2$ FL-like behavior. The values of
$\gamma$ were estimated at $T^2 =$ 0.35 K$^2$ ($T \approx$ 0.6 K)
and decrease with the increasing magnetic field (Fig.
\ref{gammaA}(a)) in a manner similar to what was observed in
YbRh$_2$Si$_2$ \cite{geg02a} and other materials. Although the
data set is sparse, it is worth noting that an approximately 50-55
kOe shift down of the data for $H \parallel c$ (Fig.
\ref{gammaA}(a)) brings it into rough agreement with the $H
\parallel (ab)$ data and so that the two sets form a universal
curve. The $T^2$ coefficient of FL-like resistivity, $A$, also
decreases with an increase of applied field (Fig.
\ref{gammaA}(b)). The shift required to have the $A$ data for the
two $H$ orientations on the same curve is 30-25 kOe.  The field
dependence of the Kadowaki-Woods ratio, $A/\gamma^2$ \cite{kad86a}
is presented in Fig. \ref{gammaA}(c). Many of the $\gamma$ and $A$
values were reckoned for the same magnetic field. In some cases
when additional values of $A$ were available a straightforward
interpolation of $\gamma(T)$ was used. Although more data points
may be required to clarify these trends, several features are seen
in the Fig. \ref{gammaA}: the obtained values of $A/\gamma^2$ are
of the same order of magnitude, but several times higher than
$\sim$ 1$\times$10$^{-5}$ $\mu \Omega$ cm/(mJ/mol K)$^2$ obtained
in \cite{kad86a} and corroborated by the larger set of data in
\cite{con94a,tsu03a}; for $H \parallel (ab)$ the Kadowaki-Woods
ratio decreases with an increase of the field (not enough data is
available for $H \parallel c$). Though magnetotransport
measurements down to lower temperatures will allow for the
estimate of $A$ in a wider temperature range and may refine our
$A/\gamma^2$ data, both of the features seen in Fig.
\ref{gammaA}(c) were observed in YbRh$_2$Si$_2$ \cite{cus03a} and
apparently are common for materials where NFL behavior can be
induced by magnetic field. In addition, a larger value of the
Kadowaki-Woods ratio is anticipated theoretically in the close
vicinity of a magnetic instability \cite{tak96a}, in agreement
with our experimental data, whereas constant ({\it i.e.}
field-independent in our case) Kadowaki-Woods ratio is expected
only in the local critical regime \cite{con01a}.

\begin{figure}
\begin{center}
\includegraphics[angle=0,width=100mm]{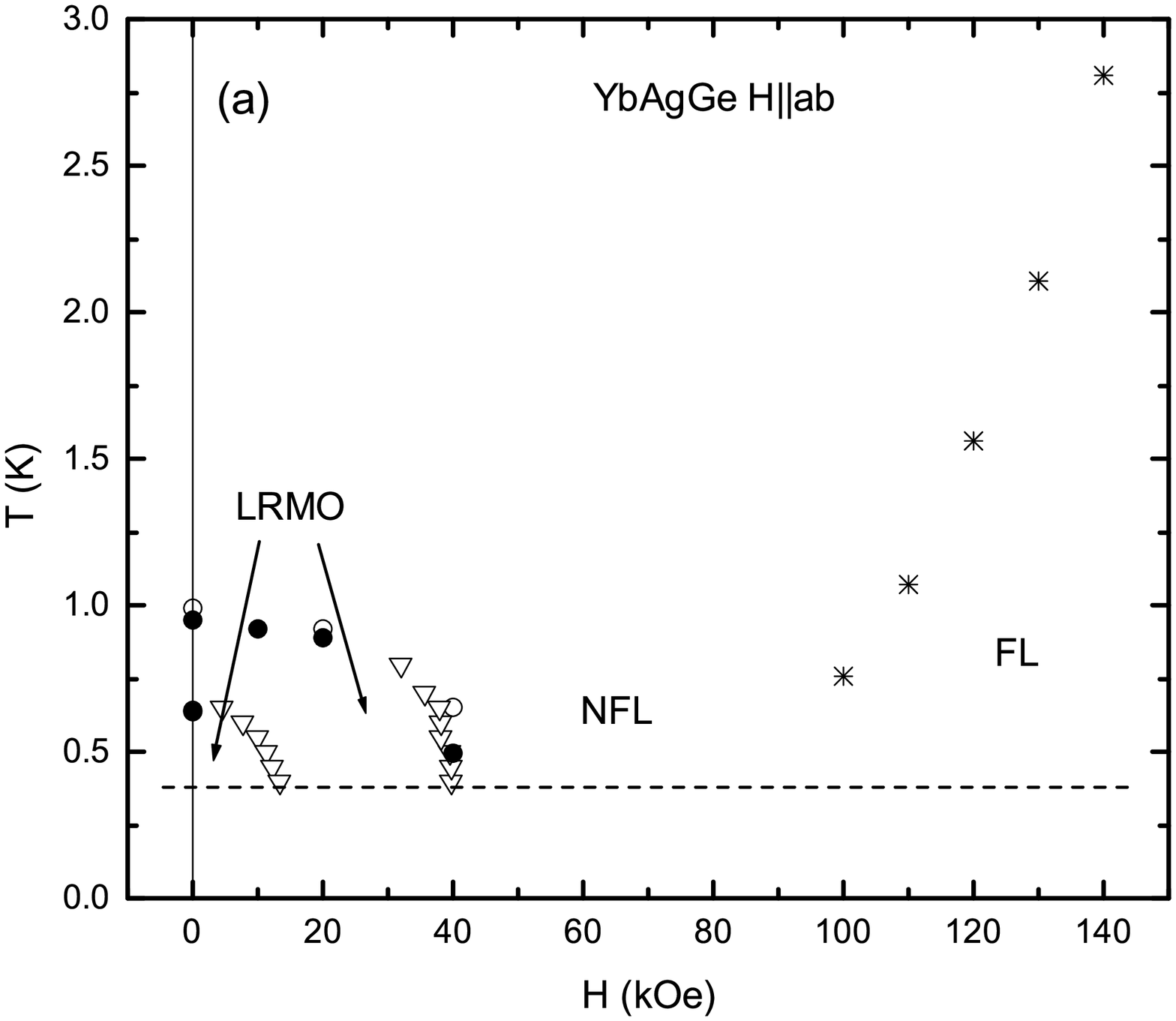}
\includegraphics[angle=0,width=100mm]{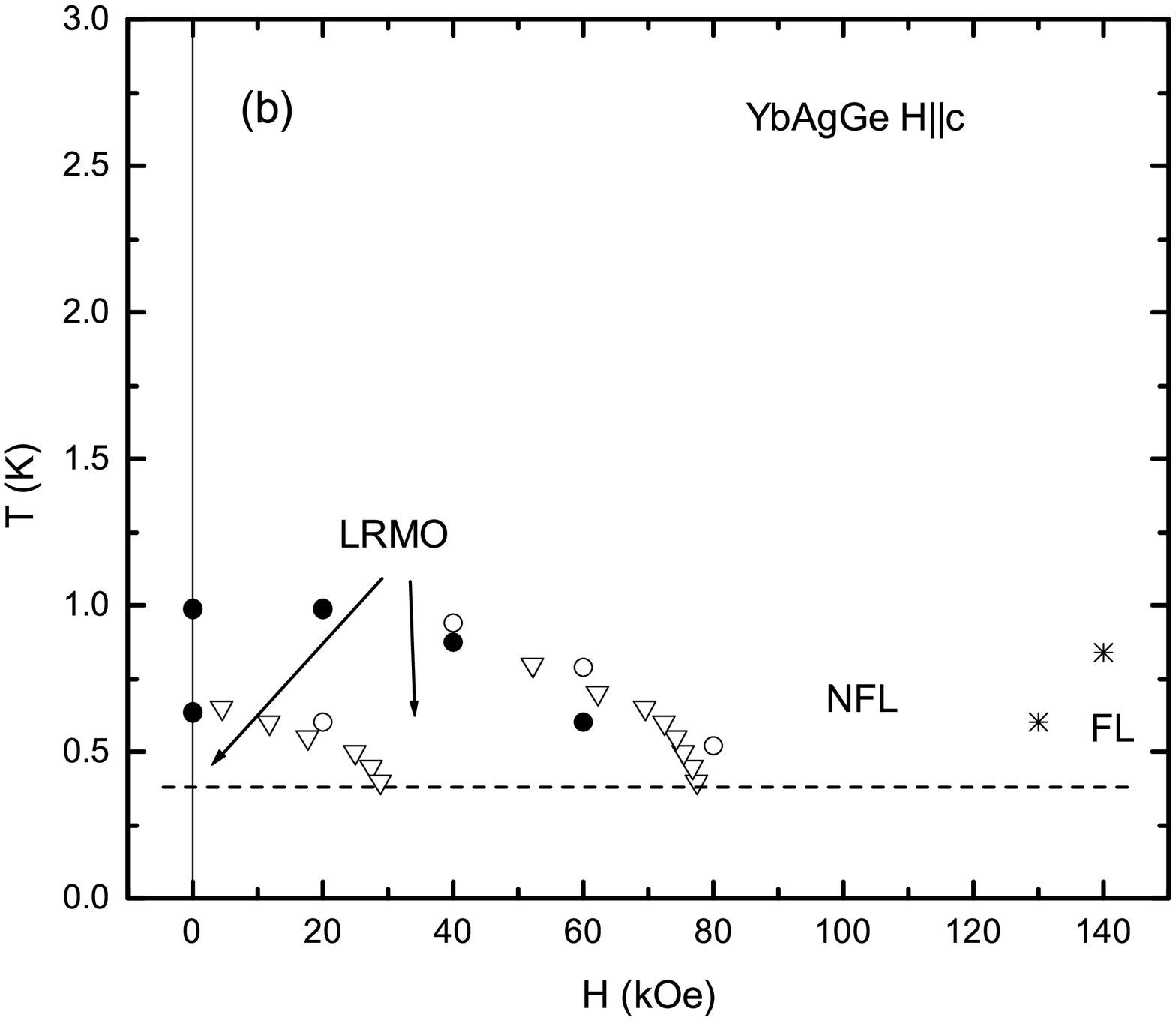}
\end{center}
\caption{Tentative $T - H$ phase diagram for (a)$H
\parallel (ab)$; (b)$H \parallel c$. Long range
magnetic order (LRMO), NFL and FL regions are marked on the phase
diagram. Symbols: Filled circles - from $C_p\mid_H$ measurements,
open circles - from $\rho(T)\mid_H$, open triangles - from
$\rho(H)\mid_T$, asterisks - temperatures below which $\Delta \rho
\propto AT^2$ in $\rho(T)\mid_H$ data (coherence temperature,
$T_{coh}$). Dashed line - low temperature limit of our
measurements, vertical line marks $H = 0$.}\label{PD}
\end{figure}

Finally, based on the thermodynamic and transport data down to
$\sim$ 0.4 K and up to 140 kOe, we can construct tentative $T - H$
phase diagrams for the two orientations of the applied magnetic
field (Fig. \ref{PD}). Both phase diagrams are very similar.
Initially increaseing magnetic field drives first the lower and
then the higher magnetic transitions to zero. With further
increase in field signatures of the NFL behavior appear in the
temperature dependent resistivity ($\Delta \rho \propto T$) and
heat capacity ($C_{magn}/T \propto - \ln T$) and at our highest
applied field values FL-like low temperature resistivity ($\Delta
\rho \propto T^2$) ({\it i.e.} the coherence line
\cite{con01a,con89a} on the $T - H$ phase diagram) is observed.
Although the current lack of data below $\sim$ 0.4 K impairs our
ability to fully delineate the critical field that corresponds to
$T =$ 0 QCP, a rather crude assessment of the data (Fig. \ref{PD})
suggests $H_c^{ab} \approx$ 45-70 kOe, $H_c^c \approx$ 80-110 kOe.

\section{Conclusions}
We presented results that allow for the classification of YbAgGe
as a new heavy fermion material with magnetic field induced NFL
behavior (critical fields are $H_c^{ab} \approx$ 45-70 kOe, $H_c^c
\approx$ 80-110 kOe). Although its critical fields are somewhat
higher than found for the extensively studied YbRh$_2$Si$_2$, they
are still within the range accessible by many groups. It should be
mentioned that only very few {\it stoichiometric} compounds are
known to demonstrate this type of behavior, making YbAgGe an
important and interesting addition to the family of strongly
correlated materials. Results of this work can serve as a road map
for further studies, delineating further experimental courses:
macroscopic (magnetization) and microscopic (neutron diffraction,
$\mu$SR, Moessbauer spectroscopy) measurements at low temperatures
and in applied field are desirable to clarify the nature of the
magnetically ordered states in YbAgGe and their evolution in
field; lower temperature ($T <$ 0.4 K), detailed thermodynamic and
transport measurements in the vicinity of the field-induced QCP
would be very helpful for the understanding of the physics of
field induced NFL behavior and as a point of comparison with a
number of existing theories \cite{ste01a,col01a,pog03a,con03a} and
with other materials with similar behavior. In addition, as is
often the case for materials close to QCP, pressure and doping
study may have a great potential in fine tuning of the ground
state properties of YbAgGe.

\begin{acknowledgments}
Ames Laboratory is operated for the U.S. Department of Energy by
Iowa State University under Contract No. W-7405-Eng.-82. This work
was supported by the Director for Energy Research, Office of Basic
Energy Sciences. S.L.B. thanks M. A. Continentino for useful
discussions.
\end{acknowledgments}

\clearpage

\clearpage

\clearpage

\clearpage

\clearpage

\clearpage

\clearpage

\clearpage

\clearpage

\clearpage

\end{document}